\documentclass[fleqn,usenatbib]{mnras}

\usepackage{newtxtext,newtxmath}

\usepackage[T1]{fontenc}
\usepackage{ae,aecompl}

\usepackage{graphicx}   
\usepackage{amsmath}    
\usepackage{amssymb}    

\usepackage{textcomp}
\usepackage{gensymb}
\usepackage{hyperref}
\usepackage{multirow}
\usepackage{setspace}
\usepackage[normalem]{ulem}
\useunder{\uline}{\ul}{}

\usepackage{rotating}
\usepackage{adjustbox}
\usepackage[utf8]{inputenc} 

\usepackage{subcaption}
\usepackage{pdflscape}	

\title[LSST photo-z testbed]{Deep $ugrizY$ Imaging and DEEP2/3 Spectroscopy: A Photometric Redshift Testbed for LSST and Public Release of Data from the DEEP3 Galaxy Redshift Survey}

\author[R. Zhou et al.]{Rongpu Zhou,$^{1,2}$\thanks{E-mail: rongpu.zhou@pitt.edu}
Michael C. Cooper,$^{3}$
Jeffrey A. Newman,$^{1,2}$
\newauthor
Matthew L. N. Ashby,$^{4}$
James Aird,$^{5}$
Christopher J. Conselice,$^{6}$
Marc Davis,$^{7,8}$
\newauthor
Aaron A. Dutton,$^{9}$
S. M. Faber,$^{10}$
Jerome J. Fang,$^{11}$
G. G. Fazio,$^{4}$
\newauthor
Puragra Guhathakurta,$^{10}$
Dale Kocevski,$^{12}$
David C. Koo,$^{10}$
Kirpal Nandra,$^{13}$
\newauthor
Andrew C. Phillips,$^{10}$
David J. Rosario,$^{14}$
Edward F. Schlafly,$^{15}$
Jonathan R. Trump,$^{16}$
\newauthor
Benjamin Weiner,$^{17}$
Christopher N. A. Willmer$^{18}$
and Renbin Yan$^{19}$
\\\\
$^{1}$Department of Physics and Astronomy, University of Pittsburgh, 3941 O'Hara Street, Pittsburgh, PA 15260, USA\\
$^{2}$Pittsburgh Particle physics, Astrophysics, and Cosmology Center (PITT PACC)\\
$^{3}$Center for Cosmology, Department of Physics and Astronomy, 4129 Reines Hall, University of California, Irvine, CA 92697, USA\\
$^{4}$Center for Astrophysics | Harvard \& Smithsonian, 60 Garden Street, Cambridge, MA 02138, USA\\
$^{5}$University of Leicester, University Road, Leicester LE2 3RJ, UK\\
$^{6}$Centre for Astronomy and Particle Theory, University of Nottingham, Nottingham NG7 2RD, UK\\
$^{7}$Department of Astronomy, University of California, Berkeley, CA 94720, USA\\
$^{8}$Department of Physics, University of California, Berkeley, CA 94720, USA\\
$^{9}$New York University Abu Dhabi, PO Box 129188, Abu Dhabi, United Arab Emirates\\
$^{10}$UCO/Lick Observatory and Department of Astronomy and Astrophysics, University of California, Santa Cruz, 1156 High Street, Santa Cruz, CA 95064, USA\\
$^{11}$Astronomy Department, Orange Coast College, Costa Mesa, CA 92626, USA\\
$^{12}$Department of Physics and Astronomy, Colby College, Waterville, ME 04961, USA\\
$^{13}$Max-Planck-Institut für Extraterrestrische Physik, Giessenbachstra{\ss}e, D-85748 Garching, Germany\\
$^{14}$Centre for Extragalactic Astronomy, Department of Physics, Durham University, Durham, DH1 3LE, UK\\
$^{15}$Lawrence Berkeley National Laboratory, One Cyclotron Road, Berkeley, CA 94720, USA\\
$^{16}$Department of Physics, University of Connecticut, 2152 Hillside Road, Unit 3046, Storrs, CT 06269, USA\\
$^{17}$MMT/Steward Observatory, University of Arizona, 933 North Cherry Avenue, Tucson, AZ 85721, USA\\
$^{18}$Steward Observatory, University of Arizona, 933 North Cherry Avenue, Tucson, AZ 85721, USA\\
$^{19}$Department of Physics and Astronomy, University of Kentucky, 505 Rose Street, Lexington, KY 40506-0055, USA
}

\begin{document}
\label{firstpage}
\pagerange{\pageref{firstpage}--\pageref{lastpage}}
\maketitle

\begin{abstract}
We present catalogs of calibrated photometry and spectroscopic redshifts in the Extended Groth Strip, intended for studies of photometric redshifts (photo-$z$'s). The data includes $ugriz$ photometry from CFHTLS and $Y$-band photometry from the Subaru Suprime camera, as well as spectroscopic redshifts from the DEEP2, DEEP3 and 3D-HST surveys.   These catalogs incorporate corrections to produce effectively matched-aperture photometry across all bands, based upon object size information available in the catalog and Moffat profile point spread function fits. We test this catalog with a simple machine learning-based photometric redshift algorithm based upon Random Forest regression, and find that the corrected aperture photometry leads to significant improvement in photo-$z$ accuracy compared to the original SExtractor catalogs from CFHTLS and Subaru. The deep $ugrizY$ photometry and spectroscopic redshifts are well-suited for empirical tests of photometric redshift algorithms for LSST. The resulting catalogs are publicly available at \url{http://d-scholarship.pitt.edu/36064/}.  We include a basic summary of the strategy of the DEEP3 Galaxy Redshift Survey to accompany the recent public release of DEEP3 data.
\end{abstract}

\begin{keywords}
galaxies: distances and redshifts -- catalogues -- surveys
\end{keywords}



\section{Introduction}
\label{sec:introduction}
Redshift is a crucial observable in the study of galaxies and cosmology. Spectroscopic redshifts are accurate, but the observations required are much more expensive than photometric measurements. Modern imaging surveys can measure the photometry of a huge number of objects very efficiently, but only a very small fraction will have observed spectra. For such surveys, redshifts must be estimated from broad-band photometry, and the large number of photometric redshift (photo-$z$) measurements compensates for their inaccuracy. The availability of large imaging datasets has made photometric redshift estimates an increasingly important component of modern extragalactic astronomy and cosmology studies.

The Large Synoptic Survey Telescope \citep{LSST09, Ivezic09} will rely on photometric redshifts to achieve many of its science goals. For ten years, LSST will survey the sky in six filters to a depth unprecedented over such a wide area. The resulting dataset should provide important clues to the nature of dark matter and dark energy, detailed information on the structure of the Milky Way, a census of near-earth objects in the Solar System, and a wealth of information on variable and transient phenomena. In this paper, we present catalogs with robust spectroscopic redshift measurements and well-calibrated photometry in the Extended Groth Strip (EGS) with filter coverage and depths similar to the LSST $ugrizy$ system. The LSST Science Requirements Document\footnote{\url{www.lsst.org/scientists/publications/science-requirements-document}} specifies that for galaxies with $i < 25$ the LSST data should be capable of delivering a root mean square (RMS) error in redshift smaller than $0.02(1+z)$ with a rate of  $>3\sigma$ outliers below 10\%. The dataset we have assembled will be useful for assessing if current photometric redshift algorithms can meet these requirements, and for improving them if not.


A previous paper, \citet{Matthews13}, matched redshifts from the DEEP2 Galaxy Redshift Survey \citep{Newman13} to photometry from the Canada-France-Hawaii Telescope Legacy Survey (CFHTLS \citealt{Hudelot12}) and the Sloan Digital Sky Survey (SDSS \citealt{Gunn98,Alam15}). This works builds on that effort by adding DEEP3 \citep{Cooper11, Cooper12} and 3D-HST \citep{Brammer12,Momcheva16} redshifts and $Y$-band photometry, and using Pan-STARRS \citep{Chambers2016, Magnier16} instead of SDSS for photometric calibration. We also have developed a method for calculating corrected aperture photometry from the CFHTLS catalogs, and we perform tests with a simple photometric redshift algorithm to demonstrate the superiority of this photometry for measuring galaxy colors.

The structure of this paper is as follows. Section \ref{sec:datasets} describes the datasets that we used to produce the final catalogs. We use spectroscopic redshifts from the DEEP2 and DEEP3 surveys, as well as grism redshifts from 3D-HST. The photometry in the $ugriz$ bands is from CFHTLS. Additionally, $Y$-band imaging was obtained from SuprimeCam at the Subaru telescope \citep{Miyazaki02}; photometry based on these images was derived using SExtractor \citep{Bertin96}. In section \ref{sec:astrometry} we describe the methods used to bring the CFHTLS, Subaru $Y$-band, and Pan-STARRS1 catalogs to a common astrometric system, based on those employed by \citet{Matthews13}. We describe our photometric zero-point calibration methods in section \ref{sec:photometry} and the techniques used to produce corrected aperture photometry in section \ref{sec:MAG_APERCOR}. In section \ref{sec:combined} we describe the resulting matched catalogs, which are being released in concert with this paper. In section \ref{sec:photo-z} we present tests of these catalogs using photometric redshifts measured via Random Forest regression. We provide a summary in section \ref{sec:summary}.

\section{Datasets}
\label{sec:datasets}

In this section, we describe the spectroscopic and imaging datasets used to construct the catalogs presented in this paper.

\subsection{Spectroscopy}

The first spectroscopic sample included in our catalogs comes from the DEEP2 Galaxy Redshift Survey, which is a magnitude-limited spectroscopic survey performed using the DEIMOS spectrograph at the Keck 2 telescope. Galaxy spectra were observed in four fields, with targets lying in the magnitude range $R_{AB}<24.1$. Field 1 (corresponding to the EGS) applied no redshift pre-selection, though objects expected to be at higher redshift received greater weight in targeting. In the remaining 3 fields, DEEP2 targeted only objects expected to be in the redshift range of $z > 0.75$. Only Field 1 is used for this paper. Details of DEEP2 are given in \citet{Newman13}.

The second spectroscopic sample included constitutes the public data release of spectra from the DEEP3 Galaxy Redshift Survey \citep{Cooper11, Cooper12}, which was was primarily intended to enlarge the DEEP2 survey within the EGS field to take advantage of the wealth of multiwavelength information available there. This release is distributed at \url{http://deep.ps.uci.edu/deep3/home.html}.  We describe DEEP3 in more detail in appendix \ref{deep3} to accompany this data release.

We also incorporate grism redshift data from the 3D-HST survey \citep{Brammer12,Momcheva16}, which measures redshift down to $JH_{IR}=26$. The 3D-HST sample reaches higher redshifts than DEEP2 or DEEP3. The 3D-HST grism redshifts are derived using a combination of grism spectra and photometric data, and proper selection is needed to ensure a set of robust redshifts. The selection criteria used are described in section \ref{sec:photo-z}.

\subsection{Photometry in $ugriz$ bands}

For the $ugriz$ bands, we used the CFHTLS-T0007 \citep{Hudelot12} catalogs of photometry from CFHT/MegaCam. We utilize data from the CFHTLS Deep field D3 as well as the seven pointings in the Wide field W3 which overlap with DEEP2/3 and 3D-HST. The list of pointings may be found in Table \ref{tab:a0_a1_cfhtls}. The CFHTLS Wide field sample reaches $5\sigma$ depths of $u\sim24.7$, $g\sim25.4$, $r\sim24.8$, $i\sim24.3$, and $z\sim23.5$. The CFHTLS Deep data reaches $5\sigma$ depths of $u\sim27.1$, $g\sim27.5$, $r\sim27.2$, $i\sim26.9$, $i_2\sim26.6$, and $z\sim25.8$, where $i_2$ is the replacement filter for the $i$-band filter. This filter was named $y$ in the CFHTLS catalogs, but within this paper and in our catalogs we refer to this filter as $i_2$ to avoid confusion with the $y$-band in the LSST $ugrizy$ filter system. The default photometry from CFHTLS is the Kron-like elliptical aperture magnitude MAG\_AUTO. We also have calculated a set of corrected aperture magnitudes as described below, which we  designate as MAG\_APERCOR in catalogs. See section \ref{sec:MAG_APERCOR} for details of the aperture correction procedure applied.

We have utilized an internal version of the Pan-STARRS1 (PS1) catalog \citep{Chambers2016, Magnier16} to calibrate the photometric zero-points for the $griz$ and $Y$ bands. For the CFHTLS u-band we have used the Deep field photometry as the standard against which we calibrate the Wide field data, as described in section \ref{sec:u-band}. 

\subsection{$Y$-band data}
\label{sec:y-band}
In addition to the $ugriz$ bands which are included in CFHTLS, LSST will obtain data in the $y$ band. To obtain photometry of comparable depth in a similar filter, we used the $Y$-band filter available for Suprime-Cam on the Subaru telescope \citep{Miyazaki02} over the course of two nights to cover a portion of the DEEP2 EGS field. The wavelength coverage of this filter is slightly redder and narrower than the LSST $y$-band filter, but it is otherwise similar. The $Y$-band observations consist of two pointings centered on $\mathrm{RA}=14^h17^m58.2^s$, $\mathrm{Dec}=+52\degree36\arcmin4.0\arcsec$ and $\mathrm{RA}=14^h22^m28.0^s$, $\mathrm{Dec}=+53\degree24\arcmin58.0\arcsec$, with exposure times of 234 min and 9 min, respectively. The unequal exposure times were not planned, but rather a result of the onset of poor weather conditions. The $5\sigma$ depth of the two pointings are 25.0 and 23.4 mag, respectively, and the seeing full width at half maximum (FWHM) values were $0.662\arcsec$ and $0.632\arcsec$, respectively. A mosaic was created using the Subaru/Suprime SDFRED2 pipeline \citep{Ouchi04}. The initial astrometry for the mosaic was determined using Astrometry.net \citep{Lang10}. We then used SExtractor \citep{Bertin96} to detect sources and obtain a photometric catalog. Slightly different SExtractor parameters were used for the two pointings to account for differences in depth and seeing. The parameters are listed in \autoref{sec:SExtractor}. An initial ``guess'' of the image zero-point was used for SExtractor. We determine a more accurate zero-point later in the calibration procedure as described in section \ref{sec:photometry}. A subset of the detected sources were visually inspected to optimize the parameters, enabling us to minimize false detections and to ensure that nearby and overlapping sources are de-blended properly.

SExtractor requires a weight map for processing. To create the weight map, we set the BACK\_SIZE parameter to 16, and the resulting BACKGROUND\_RMS check image was used as the weight map (in other words, the RMS of the background evaluated over 16 pixel boxes was used as a weight map). To avoid false detections near the image boundary, objects within 15 pixels of the image boundary were not used. The astrometry was further corrected by cross-matching to SDSS (cf. section \ref{sec:astrometry}). Besides the default MAG\_AUTO photometry, we also produced aperture photometry (MAG\_APER) with aperture diameters ranging from 9 pixels to 56 pixels in 1 pixel spacing (the pixel size of SuprimeCam is $0.2\arcsec$). The MAG\_APER photometry and half-light radius were used to calculate the corrected aperture magnitudes as described in section \ref{sec:MAG_APERCOR}.

\begin{figure}
    \centering
        \includegraphics[width=\columnwidth]{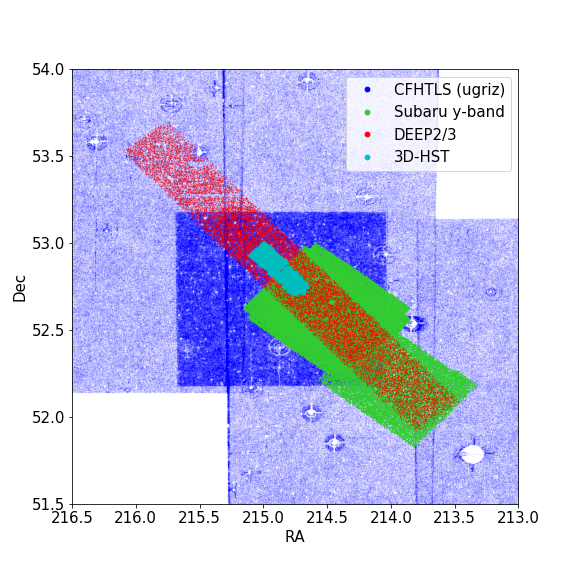}
    \caption{Sky coverage of the catalogs used in this paper (Pan-STARRS1 and SDSS are not shown as they cover the entire region). The region covered by DEEP2 and DEEP3 is shown in red, 3D-HST is in cyan, the CFHTLS imaging pointings included in our catalogs are shown in blue, and the two rectangular pointings of Subaru $Y$-band imaging are shown in green (the deeper of the two pointings partially overlaps with 3D-HST).  }
    \label{fig:sky_coverage}
\end{figure}

The sky coverage of the datasets incorporated in this work is shown in Fig. \ref{fig:sky_coverage}.

\section{Astrometric correction}
\label{sec:astrometry}

To avoid false matching between catalogs, we applied astrometric corrections to CFHTLS, the $Y$-band catalog, and PS1 to make them each match the SDSS coordinate system before cross-matching the catalogs. The astrometric offsets required varied spatially for each of these datasets.  There was no significant offset between DEEP2/3 positions (which were previously remapped to match SDSS coordinates) and SDSS. For 3D-HST, a constant RA and Dec offset were needed to match SDSS but no spatial variation in offsets was needed. 

In order to derive astrometric corrections for the CFHTLS, Subaru $Y$-band, and PS1 catalogs to match SDSS, we have applied the same methodology as described in \citet{Matthews13}.  In this paper we give only a brief outline of these techniques; we refer the reader to this prior work for details. We describe the correction of CFHTLS for sake of example. 

The correction is done separately for each pointing from CFHTLS. First we cross-match CFHTLS to SDSS with a search radius of $1.0\arcsec$. If more than one match is found, the nearest match is kept. The differences in RA and Dec ($\Delta\mathrm{RA}$ and $\Delta\mathrm{Dec}$) are calculated for every matched object. The matched objects are binned according to their RA and Dec, with a bin size of $1.2\arcmin\times1.2\arcmin$. This bin size was chosen because smaller bins did not significantly reduce the residuals and could lead to problems with over-fitting. Within each bin, the mean value of the $\Delta\mathrm{RA}$ and $\Delta\mathrm{Dec}$ are calculated using the robust Hodges-Lehmann estimator \citep{hodges1963}. For bins that have fewer than 3 objects, values from the neighboring bins are used. A $3\times3$ boxcar average is performed to smooth $\Delta\mathrm{RA}$ and $\Delta\mathrm{Dec}$, and we perform bivariate spline interpolation on the smoothed $\Delta\mathrm{RA}$ and $\Delta\mathrm{Dec}$ grid to obtain the functions $\Delta\mathrm{RA}(\mathrm{RA}, \mathrm{Dec})$ and $\Delta\mathrm{Dec}(\mathrm{RA}, \mathrm{Dec})$. For each object in the CFHTLS catalogs we then evaluate $\Delta\mathrm{RA}(\mathrm{RA}, \mathrm{Dec})$ and $\Delta\mathrm{Dec}(\mathrm{RA}, \mathrm{Dec})$ to determine the offsets at its position, and subtract them from the CFHTLS coordinates. The same method is used to correct the astrometry of PS1 and the $Y$-band catalog, with the only difference being the bin sizes used ($4\arcmin\times4\arcmin$ and $1.7\arcmin\times1.7\arcmin$, respectively, for PS1 and Subaru). Table \ref{tab:astrometry} lists the mean and standard deviation of $\Delta\mathrm{RA}$ and $\Delta\mathrm{Dec}$ for each catalog before and after these corrections.

\begin{table}
    \begin{center}
    \caption{The mean and RMS of $\mathrm{RA_{CFHTLS}} - \mathrm{RA_{SDSS}}$, $\mathrm{RA_{Subaru}} - \mathrm{RA_{SDSS}}$ and $\mathrm{RA_{PS1}} - \mathrm{RA_{SDSS}}$. The values before correction {are listed as plain text and the values after correction are in \textit{italic} font. The astrometric corrections applied are described in section \ref{sec:astrometry}.}}
    \label{tab:astrometry}
    \renewcommand{\arraystretch}{0.8}
    \begin{tabular}{|c|cc|cc|}
        \hline
        \multirow{2}{*}{Pointing} & \multicolumn{2}{ |c| }{$\mathrm{RA} - \mathrm{RA_{SDSS}}(\arcsec)$} & \multicolumn{2}{ |c| }{$\mathrm{dec} - \mathrm{dec_{SDSS}}(\arcsec)$}  \\
        & mean & $\sigma$ & mean & $\sigma$ \\ \hline
        \multirow{2}{*}{CFHTLS D3} & 0.071 & 0.303 & -0.023 & 0.180\\
        & \textit{0.003} & \textit{0.267} & \textit{0.002} & \textit{0.155}\\ \hline
        \multirow{2}{*}{CFHTLS W3-0-1} & 0.107 & 0.286 & 0.016 & 0.157\\
        & \textit{0.002} & \textit{0.257} & \textit{0.000} & \textit{0.150}\\ \hline
        \multirow{2}{*}{CFHTLS W3-1-2} & 0.058 & 0.271 & 0.042 & 0.163\\
        & \textit{0.002} & \textit{0.258} & \textit{0.001} & \textit{0.152}\\ \hline
        \multirow{2}{*}{CFHTLS W3-0-3} & 0.125 & 0.281 & -0.011 & 0.155\\
        & \textit{0.004} & \textit{0.243} & -\textit{0.001} & \textit{0.148}\\ \hline
        \multirow{2}{*}{CFHTLS W3+1-2} & 0.075 & 0.269 & -0.027 & 0.155\\
        & \textit{0.001} & \textit{0.252} & \textit{0.000} & \textit{0.147}\\ \hline
        \multirow{2}{*}{CFHTLS W3-0-2} & 0.107 & 0.284 & -0.007 & 0.158\\
        & \textit{0.001} & \textit{0.259} & \textit{0.000} & \textit{0.151}\\ \hline
        \multirow{2}{*}{CFHTLS W3+1-1} & 0.094 & 0.266 & 0.007 & 0.150\\
        & \textit{0.002} & \textit{0.243} & \textit{0.000} & \textit{0.146}\\ \hline
        \multirow{2}{*}{CFHTLS W3-1-3} & 0.033 & 0.252 & -0.003 & 0.157\\
        & \textit{0.003} & \textit{0.244} & \textit{0.000} & \textit{0.147}\\ \hline
        \multirow{2}{*}{Subaru \textit{Y}-band} & -0.042 & 0.285 & -0.165 & 0.296\\
        & \textit{-0.001} & \textit{0.259} & \textit{0.000} & \textit{0.151}  \\ \hline
        \multirow{2}{*}{PS1} & 0.020  & 0.285 & -0.022 & 0.171\\
        & \textit{-0.001} & \textit{0.264} & \textit{0.000} & \textit{0.153}\\ \hline
    \end{tabular}
    \end{center}
\end{table}

\section{Photometric zero-point calibration}
\label{sec:photometry}
The CFHTLS photometry is in the AB system but has systematic zero-point offsets that must be corrected. We also need to determine the $Y$-band zero-point. PS1 has $grizy$ photometry that is well-calibrated \citep{Magnier16}, so it is well-suited to use as a standard for improving the calibration of most bands used in this work. The calibration of CFHTLS $u$-band must be handled differently, however, since this filter is not observed by PS1. Our methods for  $u$-band calibration are described in section \ref{sec:u-band}.

\subsection{Pan-STARRS1 catalog}
The PS1 catalog contains columns corresponding to the mean flux, median flux and flux error in each band for all objects. For convenience we convert the mean flux and flux error to AB magnitude and magnitude error via standard error propagation. To eliminate false detections, we require that an object has at least three ``good'' detections (nmag\_ok$\geq$1) in the six bands. The PS1 photometry has been found to have small zero-point offsets compared to the standard AB system \citep{Scolnic15}; we have shifted the PS1 $grizy$ magnitudes by +20, +33, +24, +28, and +11 mmag ($griz$ offsets from Table 3 of \citealt{Scolnic15}; $y$-band offset from private communication from Dan Scolnic), respectively, to match to the AB system.

\subsection{Zero-point calibration of $grizY$ bands}

The filter throughputs and overall system responses vary between different telescopes even for the same nominal band, so in general the measured fluxes of the same source should differ between catalogs. However, if the filter responses are sufficiently similar and the source spectrum is nearly flat over the filter wavelength range, the brightness measured from the two telescopes should be approximately the same, as the color measured between any two instruments/filters should be zero for a flat spectrum source (by the definition of the AB system). Such flat-spectrum sources can be approximated by observed objects with zero color in the AB system; the magnitudes measured from two telescopes should be the same for these objects if all photometry is properly calibrated to AB. Based on this idea, we calculated the zero-point offset between PS1 and other photometry by performing a linear fit of magnitude difference as a function of color for stars that are found in a given pair of catalogs:
\begin{subequations}
    \begin{align}
        g_c - g_p &= a_{0, g} + a_{1, g} * (g_p - r_p), 
                \label{eq:a0_a1_g}\\
        r_c - r_p &= a_{0, r} + a_{1, r} * (r_p - i_p), \label{eq:a0_a1_r}\\
        i_c - i_p &= a_{0, i} + a_{1, i} * (i_p - z_p), 
                \label{eq:a0_a1_i}\\
        i2_c - i_p &= a_{0, i} + a_{1, i} * (i2_p - z_p), \label{eq:a0_a1_i2}\\
        z_c - z_p &= a_{0, z} + a_{1, z} * (i_p - z_p), 
                \label{eq:a0_a1_z}\\
        Y_s - y_p &= a_{0, y} + a_{1, y} * (z_p - y_p),
                \label{eq:a0_a1_y}
    \end{align}
\end{subequations}
where $a_{0, m}$ is the zero-point offset, and the subscripts c, s, and p stand for CFHTLS, Subaru and PS1, respectively. As noted previously, the variable $i2$ in equation \ref{eq:a0_a1_i2} represents the magnitude from the replacement filter for the CFHTLS $i$-band, which was slightly different from the original $i$-band filter. It is labeled as the $y$-band in CFHTLS catalogs, but we relabel it $i2$ here to avoid confusion with the Subaru $Y$-band. 

In order to perform these fits, we have cross-matched the PS1 catalog to CFHTLS and Subaru with a search radius of 1.0 arcsec. To avoid objects with large photometric errors in PS1, we require the PS1 magnitude errors to be smaller than 0.05 mag in both bands used for a given fit. Only stars that are not saturated or masked are used for calculating the offsets. For $griz$ bands, we require the ``flag'' value in the CFHTLS catalog be 0 (``star'' and ``not saturated or masked'') and the SExtractor flag in each band to be smaller than 3, providing an additional rejection of saturated objects.

To select stars for the $Y$-band, we used the star/galaxy classifier ``CLASS\_STAR'' from SExtractor, selecting those objects with $\textrm{CLASS\_STAR}>0.983$. There are a number of objects with much larger size that are misclassified as stars, and we removed them by applying a cut on the half-light radius: $r<0.44\arcsec$ for the deep pointing and $r<0.41\arcsec$ for the shallow pointing. We also removed saturated objects by requiring the SExtractor flag be smaller than 3 and applying a cut on MAG\_AUTO to reject the brightest objects, corresponding to $\mathrm{MAG\_AUTO}>17.0$ for the deep pointing and $\mathrm{MAG\_AUTO}>15.0$ for the shallow pointing.

\begin{figure*}
    \centering
    \begin{subfigure}[b]{0.33\textwidth}
        \includegraphics[width=\textwidth]{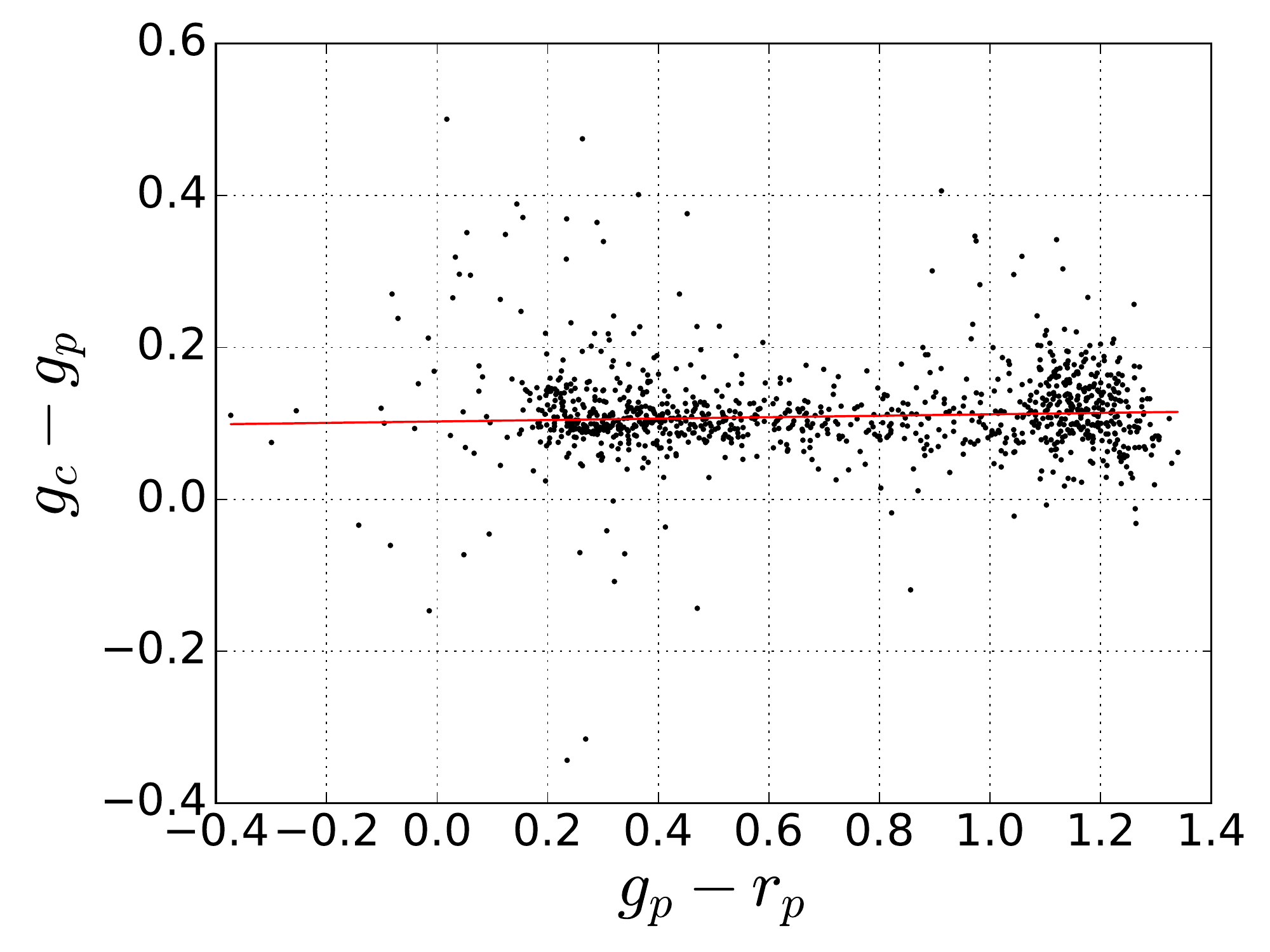}
        \caption{}
    \end{subfigure}
    \begin{subfigure}[b]{0.33\textwidth}
        \includegraphics[width=\textwidth]{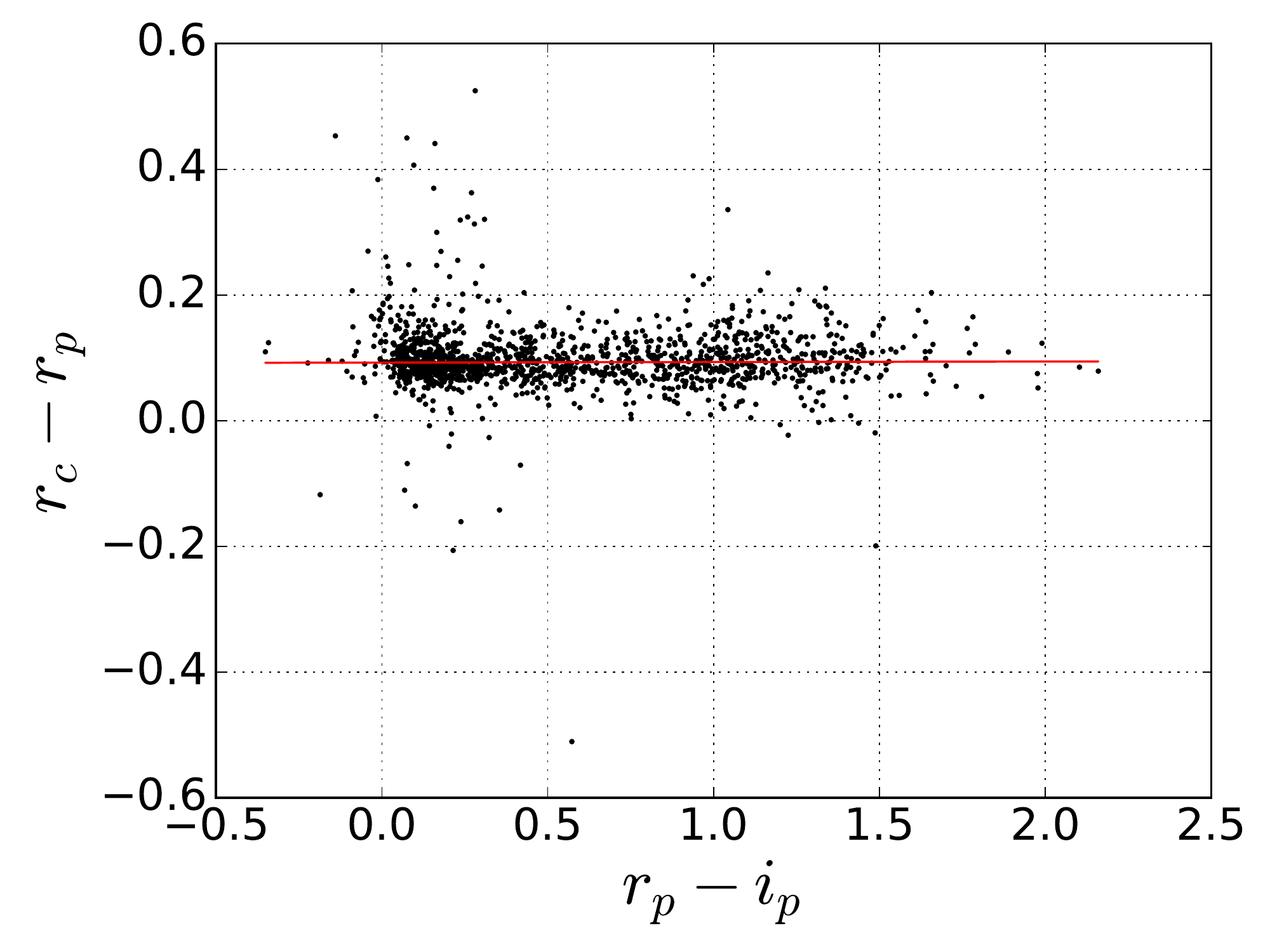}
        \caption{}
    \end{subfigure}
    \begin{subfigure}[b]{0.33\textwidth}
        \includegraphics[width=\textwidth]{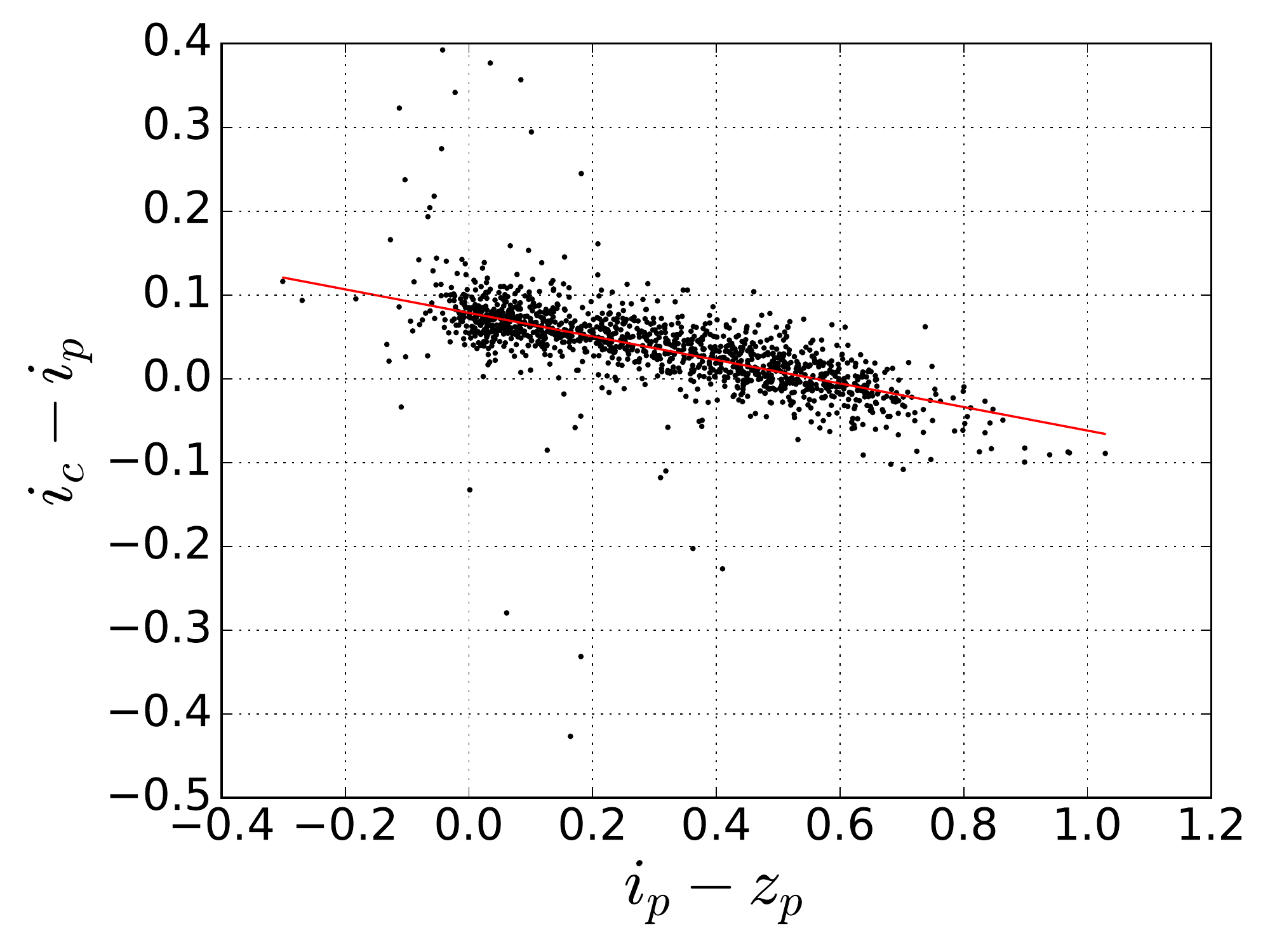}
        \caption{}
    \end{subfigure}\\
    \begin{subfigure}[b]{0.33\textwidth}
        \includegraphics[width=\textwidth]{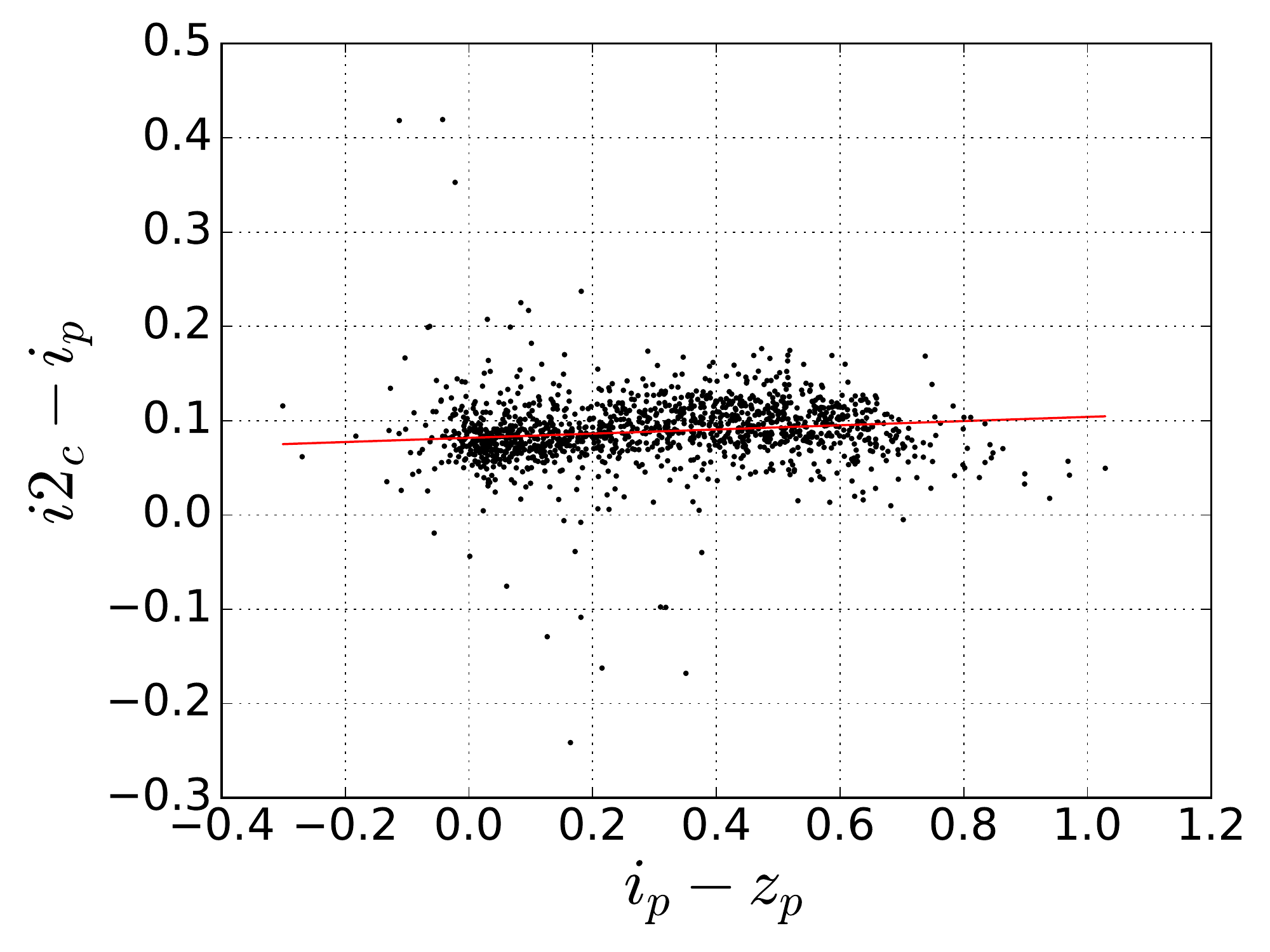}
        \caption{}
    \end{subfigure}
    \begin{subfigure}[b]{0.33\textwidth}
        \includegraphics[width=\textwidth]{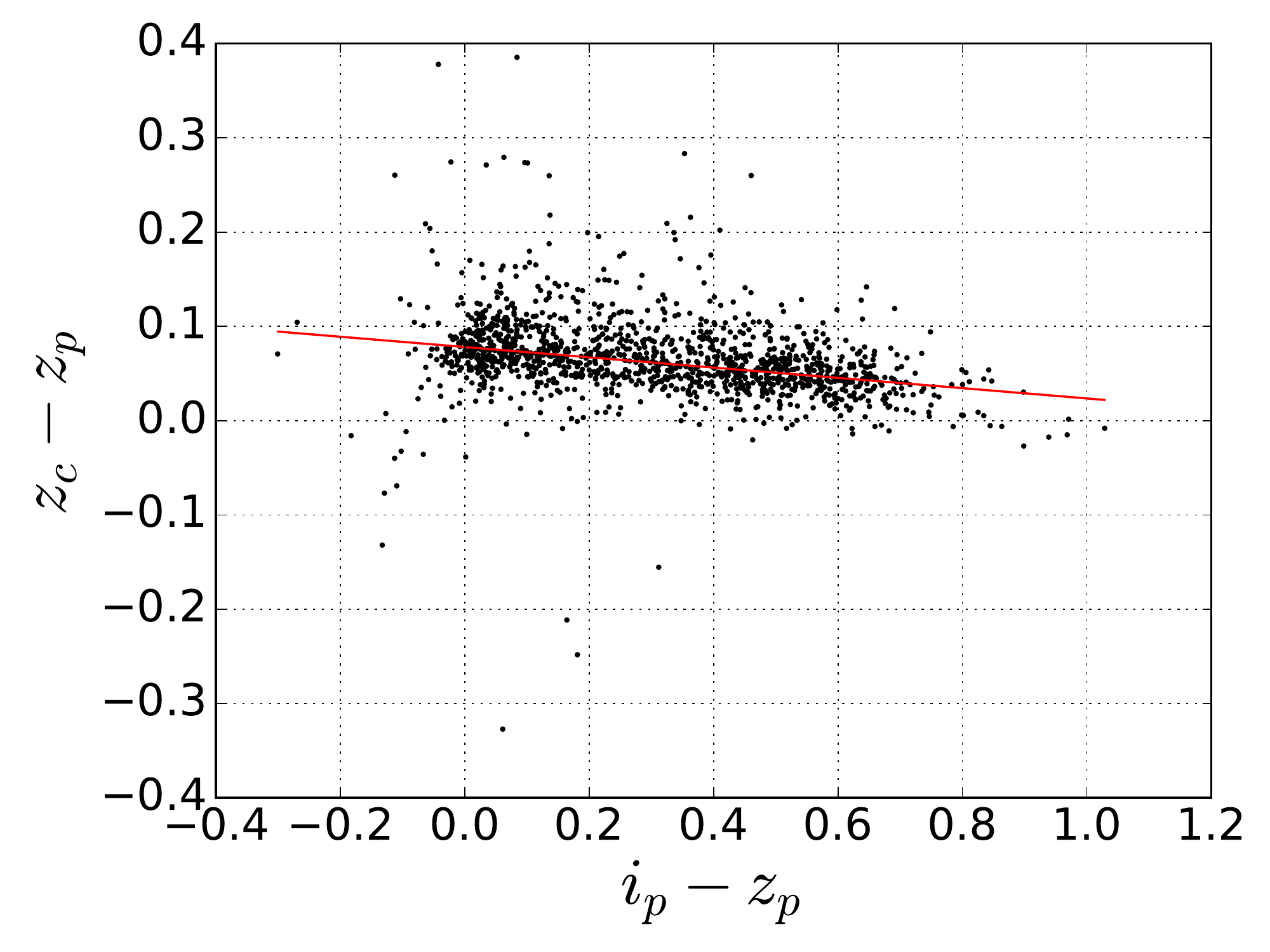}
        \caption{}
    \end{subfigure}
    \begin{subfigure}[b]{0.33\textwidth}
        \includegraphics[width=\textwidth]{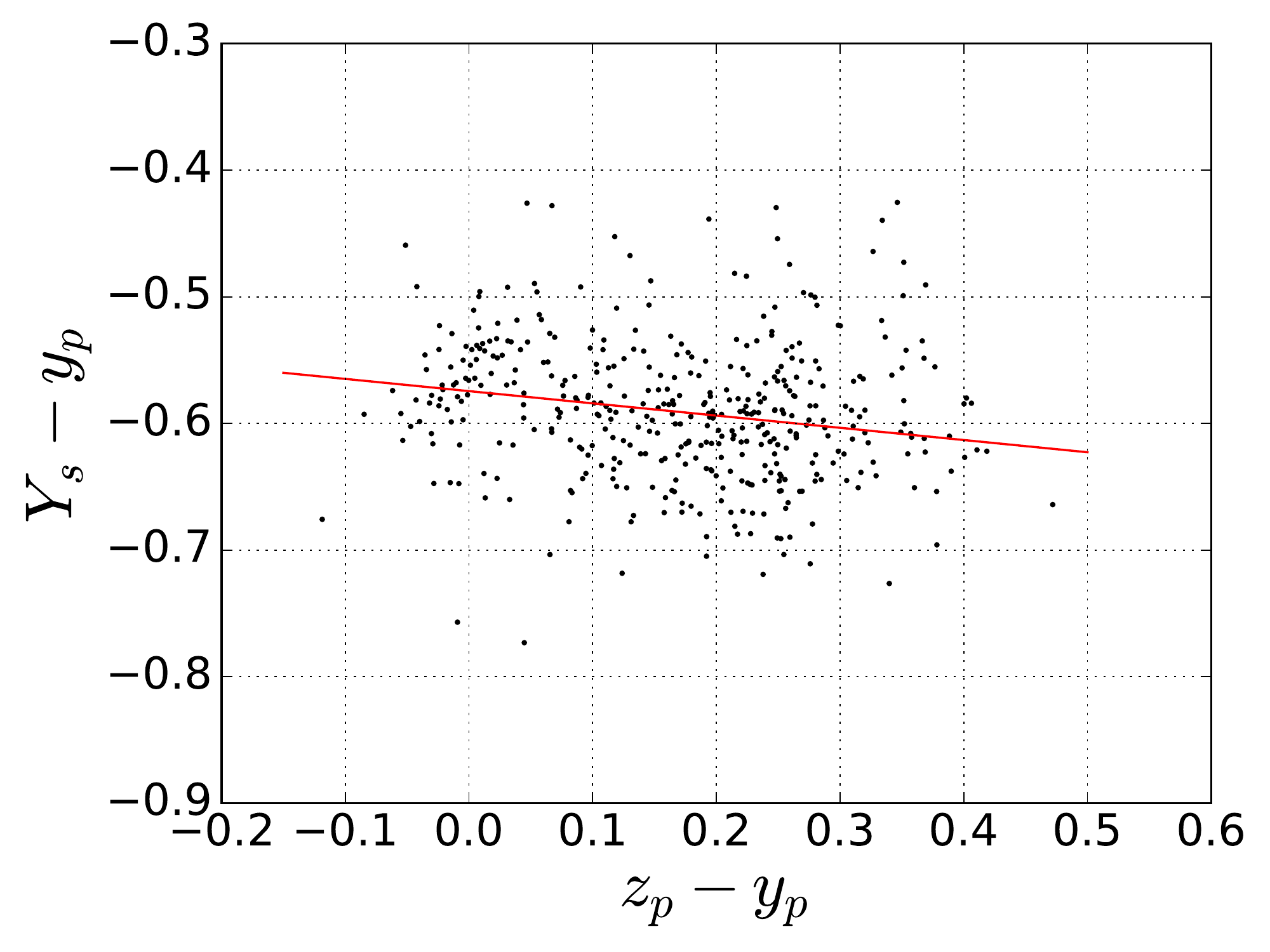}
        \caption{}
    \end{subfigure}    
    \caption{Panels (a-e) show difference in magnitude between CFHTLS Deep field D3 (subscript c) and PS1 (subscript p) plotted as a function of color. (f) shows the same plot for Subaru $Y$-band (subscript s) from the deep pointing. Only stars are used. The red lines are the linear fits described by equations \ref{eq:a0_a1_g} to \ref{eq:a0_a1_y}. The intercepts correspond to the zero-point offsets between the two systems, and are listed in Tables \ref{tab:a0_a1_cfhtls} and \ref{tab:a0_a1_subaru}. }
    \label{fig:a0_a1}
\end{figure*}

To avoid influence from outliers, we applied robust linear fitting using the Python package ``statsmodels'' and used Huber's T as an M-estimator with the tuning constant $t = 2 \mathrm{MAD}$, where MAD is the median absolute deviation between the data and the fit. The zero-point calculation is done separately for each pointing in the CFHTLS Wide field, and separately for the two $Y$-band pointings. Fig. \ref{fig:a0_a1} shows the linear fit of equations \ref{eq:a0_a1_g} to \ref{eq:a0_a1_y} using the MAG\_AUTO photometry for the CFHTLS Deep field and the Subaru deep pointing. The coefficients from the linear fits are listed in Table \ref{tab:a0_a1_cfhtls} for CFHTLS and Table \ref{tab:a0_a1_subaru} for the Subaru $Y$-band. The $a_0$ in Table \ref{tab:a0_a1_subaru} corresponds to the offset between the initial zero-point value for the $Y$-band image and the zero-point of PS1.

\begin{table*}
    \begin{center}
    \caption{Coefficients in equations \ref{eq:a0_a1_g} to \ref{eq:a0_a1_y} for CFHTLS. The coefficient $a_0$ corresponds to the zero-point offset between CFHTLS and Pan-STARRS, and is subtracted from the CFHTLS magnitudes to obtain calibrated values.}
    \label{tab:a0_a1_cfhtls}
    \begin{tabular}{|cc|cc|cc|cc|cc|cc|}
        \hline
        \multirow{2}{*}{Pointing} & \multirow{2}{*}{Method} & \multicolumn{2}{ |c| }{\textit{g} band} & \multicolumn{2}{ |c| }{\textit{r} band}  & \multicolumn{2}{ |c| }{\textit{i} band}  & \multicolumn{2}{ |c| }{\textit{i2} band}  & \multicolumn{2}{ |c| }{\textit{z} band} \\
        & & $a_0$ & $a_1$ & $a_0$ & $a_1$ & $a_0$ & $a_1$ & $a_0$ & $a_1$ & $a_0$ & $a_1$ \\ \hline
        \multirow{2}{*}{D3} & MAG\_AUTO & 0.055 & 0.004 & 0.038 & 0.000 & 0.039 & -0.128 & 0.044 & 0.029 & 0.038 & -0.048\\
        & MAG\_APERCOR & 0.017 & 0.005 & 0.000 & 0.001 & 0.004 & -0.145 & 0.010 & 0.013 & -0.002 & -0.062\\ \hline
        \multirow{2}{*}{W3-0-1} & MAG\_AUTO & 0.074 & 0.024 & 0.042 & 0.024 & 0.023 & -0.110 &  -  &  -  & 0.035 & -0.029\\
        & MAG\_APERCOR & 0.048 & -0.007 & 0.002 & 0.012 & -0.008 & -0.143 &  -  &  -  & 0.001 & -0.064\\ \hline
        \multirow{2}{*}{W3-1-2} & MAG\_AUTO & 0.081 & 0.015 & 0.036 & 0.011 & 0.042 & -0.122 &  -  &  -  & 0.062 & -0.036\\
        & MAG\_APERCOR & 0.039 & -0.001 & 0.011 & 0.001 & 0.000 & -0.150 &  -  &  -  & 0.012 & -0.067\\ \hline
        \multirow{2}{*}{W3-0-3} & MAG\_AUTO & 0.071 & 0.006 & 0.036 & 0.005 & 0.032 & -0.128 &  -  &  -  & 0.061 & -0.036\\
        & MAG\_APERCOR & 0.033 & 0.000 & -0.004 & -0.002 & -0.001 & -0.158 &  -  &  -  & 0.006 & -0.061\\ \hline
        \multirow{2}{*}{W3+1-2} & MAG\_AUTO & 0.062 & -0.005 & 0.064 & -0.002 & 0.025 & -0.124 &  -  &  -  & 0.053 & -0.040\\
        & MAG\_APERCOR & 0.032 & -0.007 & 0.020 & -0.010 & -0.006 & -0.143 &  -  &  -  & 0.006 & -0.061\\ \hline
        \multirow{2}{*}{W3-0-2} & MAG\_AUTO & 0.053 & 0.012 & 0.067 & 0.017 & 0.030 & -0.127 &  -  &  -  & 0.060 & -0.031\\
        & MAG\_APERCOR & 0.019 & 0.010 & 0.027 & 0.001 & 0.001 & -0.149 &  -  &  -  & 0.013 & -0.068\\ \hline
        \multirow{2}{*}{W3+1-1} & MAG\_AUTO & 0.067 & 0.008 & 0.055 & 0.005 & 0.018 & -0.119 &  -  &  -  & 0.058 & -0.010\\
        & MAG\_APERCOR & 0.031 & 0.000 & 0.015 & -0.004 & -0.003 & -0.144 &  -  &  -  & 0.003 & -0.060\\ \hline
        \multirow{2}{*}{W3-1-3} & MAG\_AUTO & 0.065 & 0.000 & 0.056 & 0.001 & 0.025 & -0.112 &  -  &  -  & 0.027 & -0.031\\
        & MAG\_APERCOR & 0.028 & -0.006 & 0.015 & -0.009 & -0.013 & -0.142 &  -  &  -  & -0.016 & -0.060\\ \hline
    \end{tabular}
    \end{center}
\end{table*}

\begin{table}
    \begin{center}
    \caption{Coefficients in equation \ref{eq:a0_a1_y} for Subaru \textit{Y}-band photometry. The coefficient $a_0$ corresponds to the zero-point offset between initial zero-point value for the $Y$-band image and PS1. These offsets are subtracted from the \textit{Y}-band magnitude to obtain calibrated values.}
    \label{tab:a0_a1_subaru}
    \begin{tabular}{|cc|cc|}
        \hline
        Pointing & Method & $a_0$ & $a_1$ \\ \hline
        \multirow{2}{*}{Deep} & MAG\_AUTO & -0.584 & -0.097 \\
        & MAG\_APERCOR & -0.646 & -0.101 \\ \hline
        \multirow{2}{*}{Shallow} & MAG\_AUTO & -0.653 & -0.145 \\
        & MAG\_APERCOR & -0.695 & -0.142 \\ \hline
    \end{tabular}
    \end{center}
\end{table}

So far we have assumed that the zero-point offset is uniform in each pointing. That might not be the case, and we also tried correcting for any spatial variations of the zero-point offset. To do this, we used a fixed value of the slope $a_1$ from the previous fit, and calculated the zero-point offset $a_0$ for each matching star. For example, the $g$-band offset for the j-th object is calculated as follows:
\begin{equation}
    a_{j, 0, g} = g_{j, c} - g_{j, p} - a_{1, g}*(g_{j, p} - r_{j, p}).
\end{equation}

After obtaining the zero-point offsets for each object, we obtained the spatial variation of the zero-point offset $a_{0,m}\mathrm{(RA, Dec)}$ by fitting the zero-point offset to a 2nd order bivariate polynomial of RA and Dec. Then we obtained the calibrated magnitudes: $m' = m - a_{0,m}\mathrm{(RA, Dec)}$. To test if the spatial correction actually improves the photometry, we calculated the median absolute deviation (MAD) of $a_{0, m}$ before and after spatial zero-point correction. Here we randomly select 75\% of all objects to calculate the bivariate polynomial fit, and apply the correction on the other 25\%. We repeat this procedure many times to find the statistical distribution of the difference in MAD before and after correction. For corrections to be statistically significant, we require that MAD should be smaller after correction at least 95\% of the time. Only one pointing in CFHTLS met this requirement in one band ($z$-band). Thus we conclude that there is no significant improvement by applying spatially varying zero-point corrections, so uniform corrections were applied instead.

\subsection{Calibration of the $u$-band}
\label{sec:u-band}
Because there is no $u$-band in PS1, the zero-point calibration of CFHTLS $u$-band is done differently. We tried using SDSS $u$-band as the standard photometry, but we encountered difficulties with this approach. First, the SDSS $u$-band is significantly bluer (by $\sim$ 270\AA) than the CFHTLS $u$-band; as a result the slope $a_1$ is large and our assumptions are less valid. Secondly, there are not many stars near zero color in $u-g$, and the stars that do have colors near zero exhibit large scatter. What is worse, SDSS photometry is not exactly in the AB system. For the $u$-band, it is estimated that $u_{SDSS} = u_{AB} + 0.04 \:\mathrm{mag}$ with uncertainties at the 0.01 to 0.02 mag level\footnote{\url{http://www.sdss.org/dr12/algorithms/fluxcal/\#SDSStoAB}}. Because of these problems, we have instead assumed that the CFHTLS Deep field $u$-band is well calibrated based on the tests done for the SNLS survey \citep{Hudelot12}, and calibrate the $u$-band zero-point of Wide field pointings by requiring that their $u-g$ vs $g-r$ stellar locus matches that from the Deep field. According to \citet{Hudelot12}, the calibration accuracy is at the 2\% level in the $u$-band for the Deep field. Although this uncertainty in the absolute calibration remains, the  procedure we have followed ensures that all the pointings at least have a uniform zero-point offset from the AB system, ensuring consistent photometry for calculating photometric redshifts.

Because not all of the CFHTLS Wide pointings overlap with the CFHTLS Deep pointing, direct calibration of the u-band by cross-matching Wide and Deep objects is not feasible. Thus we resort to an indirect calibration approach.
Specifically, if all pointings are calibrated in the $u$, $g$ and $r$ bands, their $u-g$ vs $g-r$ stellar loci should be the same. Since $g$ and $r$ are already calibrated, the only shift in the stellar locus should be in the $u-g$ direction, and correspond to variations in the $u$-band zero-point. To tie the $u$-band zero-point of Wide field pointings to the Deep field, we therefore need to find the relative shift in the $u-g$ direction between the stellar loci in the Deep field and a Wide field pointing. 

\begin{figure*}
    \centering
        \includegraphics[width=.7\textwidth]{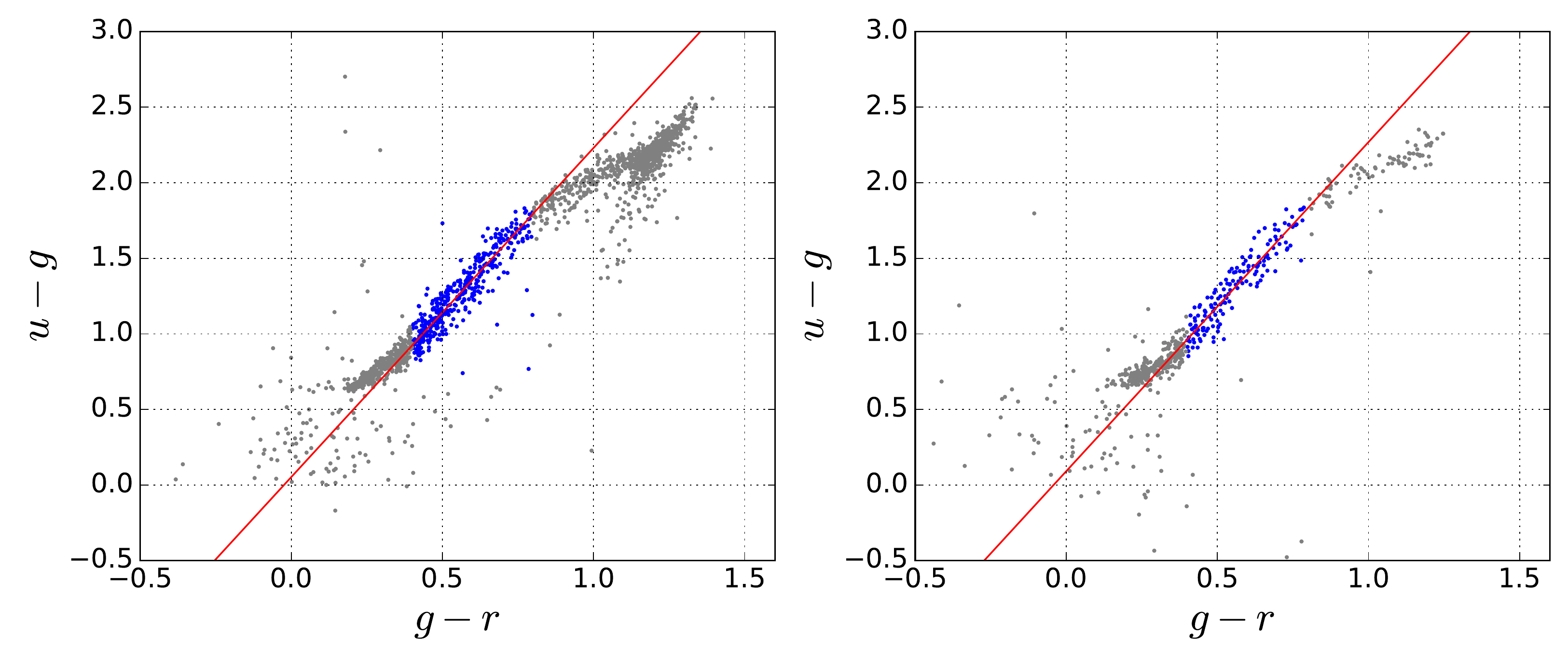}
    \caption{Stellar loci of $u-g$ vs $g-r$, using MAG\_AUTO photometry. Left panel: stellar locus and linear fit of the Deep field. The red line shows a linear fit to the points in blue. The gray points are not used for the fit. The slope of the fit is used for the Wide field pointings. Right panel: Wide field pointing W3-0-1; the red line has the same slope as in Deep field, and the difference in the intercept corresponds to the zero-point offset. }
    \label{fig:u-band}
\end{figure*}

To do this, we first selected stars in the range $0.4 < g-r < 0.8$ and $u-g>0.7$, where the stellar locus is roughly a straight line (the second cut removes outliers that are much bluer in the $g-r$ color range). The colors of the selected stars in the Deep field were fitted to a linear function. With the same color cuts, we fitted the stars in the wide field pointings with a slope fixed at the Deep field value, so that the only variable is the intercept. Fig. \ref{fig:u-band} shows the $u-g$ vs $g-r$ stellar loci and linear fits for the Deep field and one of the Wide field pointings. The differences in the intercept between the Wide field pointings and the Deep field are the $u$-band zero-point offsets, and they are listed in Table \ref{tab:u-band}.

\begin{table*}
    \caption{The \textit{u}-band zero-point offsets of the Wide field pointings relative to the Deep field. These offsets are subtracted from the Wide field \textit{u}-band magnitude to obtain calibrated values.}
    \label{tab:u-band}
    \begin{center}
    \begin{tabular}{|c|c|c|c|c|c|c|c|}
        \hline
        Pointing & W3-0-1 & W3-1-2 & W3-0-3 & W3+1-2 & W3-0-2 & W3+1-1 & W3-1-3 \\
        \hline
        MAG\_AUTO & 0.03 & 0.03 & 0.02 & 0.12 & 0.01 & 0.08 & 0.01\\
        MAG\_APERCOR & 0.06 & 0.04 & 0.02 & 0.02 & -0.02 & 0.03 & 0.00\\
        \hline
    \end{tabular}
    \end{center}
\end{table*}

\subsection{Correction for dust extinction}

The original CFHTLS $ugri(i2)z$ photometry is not corrected for Galactic extinction, nor are the PS1 magnitudes used for the photometric zero-point calibration. After zero-point calibration, we applied extinction corrections to the $ugri(i2)z$ and $Y$-band photometry. We followed the procedure described in \citet{Schlafly11}, and calculated $A_b/E(B-V)_{\mathrm{SFD}}$, where $A_b$ is the total extinction in a specific band and $E(B-V)_{\mathrm{SFD}}$ is the SFD reddening value \citep{Schlegel98}. We assumed a \citet{Fitzpatrick99} extinction law with $R_V=3.1$ and used the total transmission curves of each filter for the calculation. With $A_b/E(B-V)_{\mathrm{SFD}}$, we calculated $A_b$ using $E(B-V)_{\mathrm{SFD}}$ from the SFD dust map and applied corrections. 
Although the DEEP2/3 footprint is relatively small, there is a small spatial variation in $E(B-V)$ across the field, ranging from $0.006$ to $0.022$ with a median of $0.010$. Thus we correct for this spatial variation using the SFD map. Table \ref{tab:extinction} shows these $A_b/E(B-V)_{\mathrm{SFD}}$ values and median $A_b$ for each band.

\begin{table*}
    \caption{The values of $A_b/E(B-V)_{\mathrm{SFD}}$ in each band listed here were calculated using the procedure described in \citet{Schlafly11}. The median $A_b$ values are calculated for the set of DEEP2 and DEEP3 objects with spectroscopy.}
    \label{tab:extinction}
    \begin{center}
    \begin{tabular}{|c|c|c|c|c|c|c|c|}
        \hline
        Band & \textit{u} & \textit{g} & \textit{r} & \textit{i} & \textit{i2} & \textit{z} & \textit{y} \\
        \hline
        $A_b/E(B-V)_{\mathrm{SFD}}$ & 4.010 & 3.191 & 2.249 & 1.647 & 1.683 & 1.295 & 1.039\\
        Median $A_b$ & 0.038 & 0.031 & 0.022 & 0.016 & 0.016 & 0.012 & 0.010\\
        \hline
    \end{tabular}
    \end{center}
\end{table*}

\section{Corrected aperture photometry}
\label{sec:MAG_APERCOR}
The MAG\_AUTO from SExtractor is commonly used as the default photometry in extragalactic astronomy, and it is provided in our dataset. However it is not optimal for photometric redshift calculation for several reasons. First, it uses a relatively large aperture in order to capture most of the flux from the source, but larger apertures also lead to larger background noise. Secondly, even though a large aperture is used, it still cannot capture all the flux -- in our analysis typically $\sim$95\% of the total flux of a point source is captured by MAG\_AUTO. Thirdly, the fraction of flux captured by MAG\_AUTO might be different for objects with different sizes or images with different point spread functions (PSF's). To address these problems, we developed a method to calculate the corrected aperture photometry for both point sources and extended objects. This method utilized the aperture magnitudes at different apertures provided within the public CFHTLS catalogs, and therefore it did not require any reprocessing of the CFHTLS images. The corrected aperture magnitude is labeled ``MAG\_APERCOR'' in our catalogs. The MAG\_APERCOR photometry is calibrated the same way as MAG\_AUTO (as described in section \ref{sec:photometry}), and its zero-point offsets are listed in Table \ref{tab:a0_a1_cfhtls}, \ref{tab:a0_a1_subaru} and \ref{tab:u-band}. 

Here we summarize the techniques used for calculating ``MAG\_APERCOR''. Details can be found in Appendix \ref{sec:app_MAG_APERCOR}.  Our methods are similar to the aperture correction method described in \citet{Gawiser06}. In that work, it is assumed that all objects have a Gaussian light profile with a width calculated from the half-light radius. However, actual light profiles typically have more extended ``wings'' - i.e., more flux at large radius - than Gaussian profiles do. In our work, instead of a Gaussian profile, we have used the more flexible Moffat profile (cf. equation \ref{eq:moffat}), which has two free parameters, though we still assume that all objects have circularly symmetric light profiles that only depend on the half-light radius. This method essentially measures the flux in a small aperture ($r_0 = 0.93\arcsec$ for $ugriz$ and $r_0 = 0.9\arcsec$ for $Y$-band) and extrapolates to infinity using the Moffat profile, the parameters of which are obtained by fitting the curve of growth (the fraction of included flux as a function of aperture radius). The aperture corrections for stars and galaxies are determined slightly differently, and the $Y$-band is also treated differently since $Y$-band imaging is not  available for all objects. The steps of the aperture correction for \textit{galaxies} in band $b$ (which could be any band except $Y$) in pointing $x$ are as follows:

\begin{enumerate}
    \item Bin the objects in pointing $x$ by their $r$-band half-light radius (FLUX\_RADIUS from SExtractor);
    \item For each $r$-band radius bin, find the averaged $b$-band curve of growth and fit the Moffat profile to that curve;
    \item From the resulting best-fit parameters, obtain the correction factor $\mathrm{ApCorr} = \mathrm{Flux}(\infty)/\mathrm{Flux}(r_0)$ for each radius bin;
    \item Interpolate and extrapolate the relation between the correction factor $\mathrm{ApCorr}$ and the mean $r$-band half-light radius $R_{1/2, r}$ to obtain the continuous function $\mathrm{ApCorr_{x, b}}(R_{1/2, r})$;
    \item Use $\mathrm{ApCorr_{x, b}}(R_{1/2, r})$ and the aperture magnitude of aperture radius $r_0$ to obtain the corrected aperture magnitude.
\end{enumerate}

For stars, the procedure is the same except that they are not binned by radius, since the stars should effectively all have the same light profile set by the PSF; as a result, they are all placed in a bin together.

Although we can reduce background noise by choosing a small aperture, any errors in half-light radius will propagate into the total photometric error via the correction factor, and this can be a big problem for bands that have low S/N. For this reason, instead of using the SExtractor radius measurement in each band to assign the correction factor, we calculate the correction factor as a function of $r$-band half-light radius.  
In this way we can obtain $u$-band MAG\_APERCOR photometry even for objects with no valid radius measurement in the $u$-band. Although the absolute photometry can be affected by any $r$-band radius error, the colors are not affected as much because all bands use the same $r$-band radius for aperture correction and thus the magnitudes are all biased in the same direction. The one exception is the $Y$-band, for which we use the $Y$-band half-light radius to determine aperture corrections, as in some cases $r$ measurements may not be available or may be noisy.  The use of a matched radius makes MAG\_APERCOR well-suited for calculating photometric redshifts. A comparison of the photo-$z$ performance using MAG\_AUTO and MAG\_APERCOR is presented in section \ref{sec:photo-z}.

\section{Combined catalogs}
\label{sec:combined}
We cross-matched the CFHTLS, Subaru $Y$-band catalog and DEEP2/3 catalogs using a search radius of $1\arcsec$. CFHTLS Wide field pointings were first combined into a single catalog. For objects that appear in multiple pointings, we only kept the values from the objects that have the smallest $r$-band MAG\_APER error. Then the Wide field combined catalog was combined with the Deep field, keeping only the Deep field value if there is overlap. The combined CFHTLS catalog was then matched to the Subaru $Y$-band catalog. This final combined catalog is matched to the DEEP2/3 catalog, and all DEEP2/3 objects and columns are kept, with additional columns from CFHTLS and Subaru $Y$-band added. DEEP2/3 provides a quality flag, ``zquality''. Objects with secure redshifts can be selected by requiring $\mathrm{zquality} \geq 3$ (see \citealt{Newman13}).

Similarly, we produced a 3D-HST grism redshift catalog containing photometry from CFHTLS $ugriz$ and Subaru $Y$-band, as well as DEEP2/3 redshifts where available. To select objects with accurate grism redshifts, we require that either of the following criteria is met:
\begin{verbatim}
1.
((z_grism_u68-z_grism_l68)/(z_phot_u68-z_phot_l68)<0.1)
& ((z_grism_u68 - z_grism_l68) < 0.01) 
& (z_best_s != 0)
& (use_phot == 1)
& (z_max_grism > z_phot_l95)
& (z_max_grism < z_phot_u95)
& (z_max_grism > 0.6)
OR
2.
(z_grism_err < 0.025)
& (use_zgrism == 1),
\end{verbatim}
where all names are quantities provided in the 3D-HST catalog.  We have compared the grism redshifts selected using the above criteria with DEEP2/3 redshifts;  the normalized median absolute deviation between the spectroscopic and grism redshifts of the resulting sample is $<0.3\%$, and the fractions of objects with larger than $0.10(1+z)$ or larger than $0.02(1+z)$ redshift difference are 3\% and 11\%, respectively.
For convenience, we added a flag ``use\_zgrism1'' to the catalog, and objects that meet the above criteria are assigned the flag value 1; otherwise this flag value will be 0.

\begin{figure*}
    \centering
        \includegraphics[width=1.\textwidth]{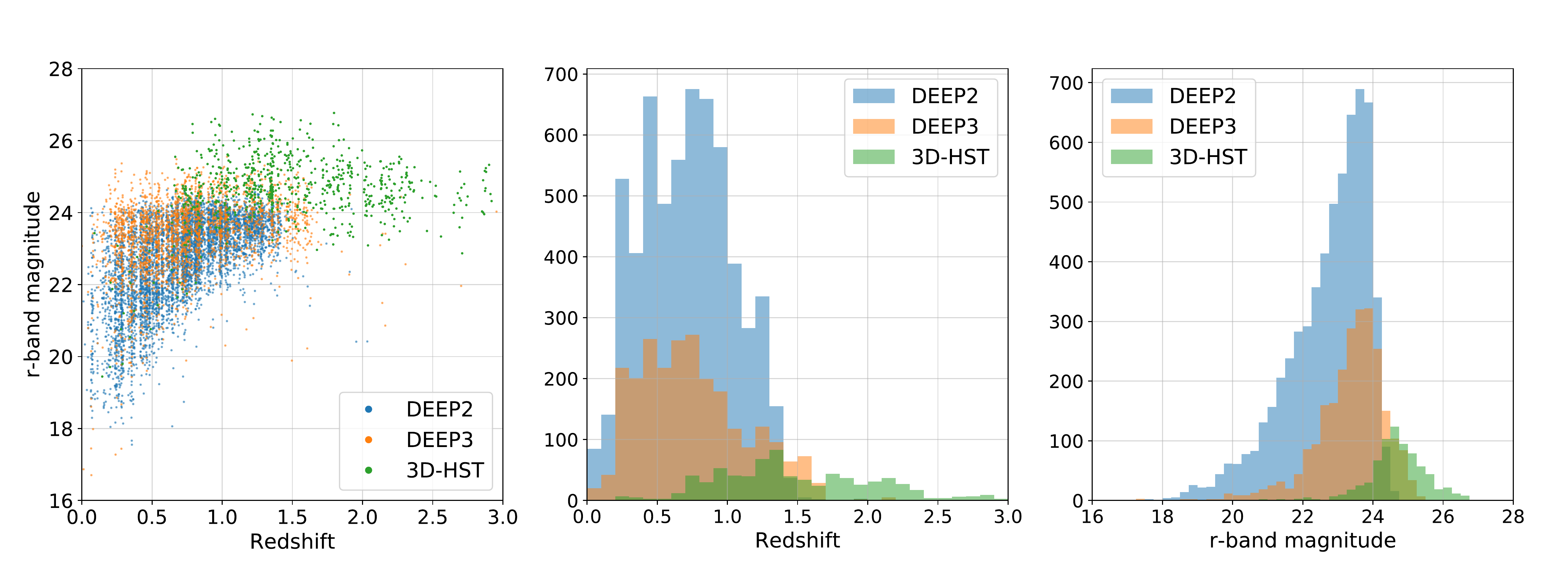}
    \caption{Left panel: $r$-band magnitude vs redshift for objects in DEEP2, DEEP3 and 3D-HST with $ugrizY$ photometry. If an object appears in both DEEP2/3 and 3D-HST, only the DEEP2/3 object is plotted. The large scale structure is clearly visible. The middle panel and the right panel show the redshift distributions and the $r$-band magnitude distributions, respectively.}
    \label{fig:redshift_and_magnitude_distribution}
\end{figure*}

Before cross-matching, the CFHTLS Deep and Wide catalogs include 603852 and 1415859 objects, respectively, and the $Y$-band catalog includes 94014 objects. The combined DEEP2/3 catalog from the aforementioned cross-matching procedures includes 8479 objects with $ugrizY$ photometry and secure DEEP2/3 redshifts, and the combined 3D-HST catalog provides an additional 741 objects with accurate grism redshifts. Fig. \ref{fig:redshift_and_magnitude_distribution} shows the distribution of $r$-band magnitude (MAG\_APERCOR) and redshift for objects with $ugrizY$ photometry and secure redshift measurements.

The columns in the catalogs are structured in the following way: the first columns listed are those from the relevant spectroscopic/grism catalog; then the columns from CFHTLS with MAG\_APERCOR and its errors are provided; and finally the $Y$-band columns are given. In the 3D-HST catalog, we also include columns of DEEP2/3 redshift and other values. Description of the DEEP2/3 columns can be found at \url{http://deep.ps.uci.edu/deep3/ztags.html} and are described in  \citet{Newman13}. Description of the 3D-HST columns can be found in Table 5 of \citet{Momcheva16}. Description of the CFHTLS columns can be found at \url{http://terapix.iap.fr/cplt/T0007/doc/T0007-docsu22.html}. The $Y$-band columns follow the same naming convention as CFHTLS. Note that the ``$y$'' variant of the CFHTLS $i$-band is relabeled $i2$ in our catalogs to limit confusion. In the catalogs, 99 indicates non-detection and -99 indicates the object/quantity is not observed. Columns of principal interest are described in Table \ref{tab:columns}.

\begin{table*}
\caption{Description of some of the principal columns included in our matched catalogs. The last three columns are DEEP2/3 values added to the 3D-HST catalog.}
\label{tab:columns}
\begin{tabular}{l|l}
    \hline
    \textbf{Column Name} & \textbf{Description} \\ \hline
    u, g, ... & MAG\_AUTO magnitude in $u$-band, $g$-band, ...\\ 
    uerr, gerr, ...  & MAG\_AUTO magnitude error in $u$-band, $g$-band, ...\\ 
    u\_apercor, g\_apercor, ... & MAG\_APERCOR magnitude in $u$-band, $g$-band, ...\\ 
    uerr\_aper, gerr\_aper, ... & MAG\_APERCOR magnitude error from image noise in $u$-band, $g$-band, ...\\ 
    uerr\_apercor, gerr\_apercor, ... & MAG\_APERCOR magnitude error from correction uncertainty in $u$-band, $g$-band, ...\\ 
    r\_radius\_arcsec & \textit{r}-band half-light radius in arcsec\\ 
    y\_radius\_arcsec & \textit{Y}-band half-light radius in arcsec\\ 
    cfhtls\_source & source of the \textit{ugri(i2)z} photometry: 0 = Deep field; 1 = Wide field; -99 = not observed\\ 
    subaru\_source & source of the \textit{Y}-band photometry: 0 = deep pointing; 1 = shallow pointing; -99 = not observed\\
    ra\_deep2, dec\_deep2 & right ascension and declination from DEEP2/3\\
    ra\_cfhtls, dec\_cfhtls & right ascension and declination from CFHTLS after astrometric correction\\
    ra\_subaru, dec\_subaru & right ascension and declination from the Subaru Y-band data after astrometric correction\\
    sfd\_ebv & E(B-V) from Schlegel, Finkbeiner, and Davis (1998) dust map\\ \hline
    	zhelio & (In DEEP2/3 catalog) DEEP2/3 heliocentric redshift\\
    zquality & (In DEEP2/3 catalog)	DEEP2/3 redshift quality flag\\ \hline
    use\_zgrism1 & (In 3D-HST catalog) our grism redshift quality flag: 0 = less accurate; 1 = accurate\\
	z\_max\_grism & (In 3D-HST catalog) 3D-HST grism redshift\\
    z\_deep2 & (In 3D-HST catalog) DEEP2/3 heliocentric redshift\\
    z\_err\_deep2 & (In 3D-HST catalog) DEEP2/3 redshift error\\
    zquality\_deep2 & (In 3D-HST catalog) DEEP2/3 redshift quality flag\\ \hline
\end{tabular}
\end{table*}

MAG\_APERCOR has two sources of error: image noise and uncertainty in aperture correction. We note that errors in colors cannot be obtained by simply adding up the two kinds of errors in quadrature due to covariances between how magnitudes were determined in each band; color errors will be smaller than one would expect if measurements in each filter were assumed to be independent. More details of how to use the errors in MAG\_APERCOR can be found in Appendix \ref{sec:app_MAG_APERCOR}.

We also provide the photometry-only catalogs of CFHTLS Wide, CFHTLS Deep and $Y$-band. These catalogs contain calibrated MAG\_AUTO and MAG\_APERCOR photometry, but are not matched to any other dataset.

\section{Photometric Redshift Tests}
\label{sec:photo-z}

In this section, we describe the photo-$z$ tests performed on the catalogs. In general, there are two classes of method for calculating the photometric redshifts. One is the template-fitting method, in which the redshift is obtained from the best fit to the photometry (in the chi-squared sense) determined using known template SEDs. The other is the empirical method, in which a dataset with spectroscopic redshifts is used to train an empirical relation between photometry and redshift (typically via machine learning algorithms), and the empirical relation is then applied to new photometric data to estimate the redshift. Here we use a machine learning algorithm called random forest regression \citep{Breiman2001} which is included in the Python package \textit{Scikit-learn} \citep{scikit-learn}. Random forest is an ensemble learning method based on decision trees. A simple decision tree is trained by minimizing the sum of squared errors, and it tends to fit the noise in the data (i.e. over-fitting). The over-fitting results in reduced accuracy when the algorithm applied to new data. Random forest addresses this problem in two ways. First, a large number of new samples are created by bootstrapping the original training sample, and separate decision trees are trained using each sample.  Secondly, instead of all the features (colors in our case), a random subset of the features may be used at each tree split to reduce the correlation between the trees. Although over-fitting can  occur in individual trees, the effect is reduced by using subsets of features and averaged out by combining the predictions from all the trees. In our analysis using a subset of features did not significantly improve the results, and thus all available features were used at each split.

Both DEEP2/3 and 3D-HST data were employed to train and assess the performance of the algorithm. The selection of DEEP2/3 and 3D-HST redshifts is described in section \ref{sec:combined}. For objects that appear in both DEEP2/3 and 3D-HST, the DEEP2/3 redshift values are used. To avoid training and testing on the same dataset, we applied the K-fold cross-validation method: the dataset is first randomly divided into 5 subsets. Then one subset is selected as the testing set and the other 4 subsets are combined as a training set for optimizing the random forest, and this procedure is repeated 5 times so that the entire dataset has been used as the testing set in the end. The estimated photometric redshift derived for a given object when it was in the testing set is then compared with the spectroscopic/grism redshift (from now on simply spectroscopic redshift or $z_{\mathrm{spec}}$ for convenience) and the redshift difference $\Delta z = z_{phot} - z_{spec}$ is calculated. Two quantities are used to evaluate the photo-$z$ performance here: the normalized median absolute deviation $\sigma_{\mathrm{NMAD}} = 1.48 \:\mathrm{MAD}$, where $\mathrm{MAD} = \mathrm{median}(|\Delta z|/(1 + z_{spec}))$, and the outlier fraction $\eta$ which is defined as the fraction of objects with $|\Delta z| > 0.15/(1+z_{\mathrm{spec}})$.

For consistent S/N in the photometry, the CFHTLS Wide field and Deep field are tested separately, and in both cases the $Y$-band photometry from both the deep and shallow pointing are used. Valid photometry in all six bands ($ugrizY$) is required. We have tested the photometric redshift performance for both MAG\_AUTO and MAG\_APERCOR photometry. The five colors $u-g$, $g-r$, $r-i$, $i-z$, $z-y$ and $i$-band magnitude are used as the input. 

\begin{figure*}
    \centering
        \includegraphics[width=.8\textwidth]{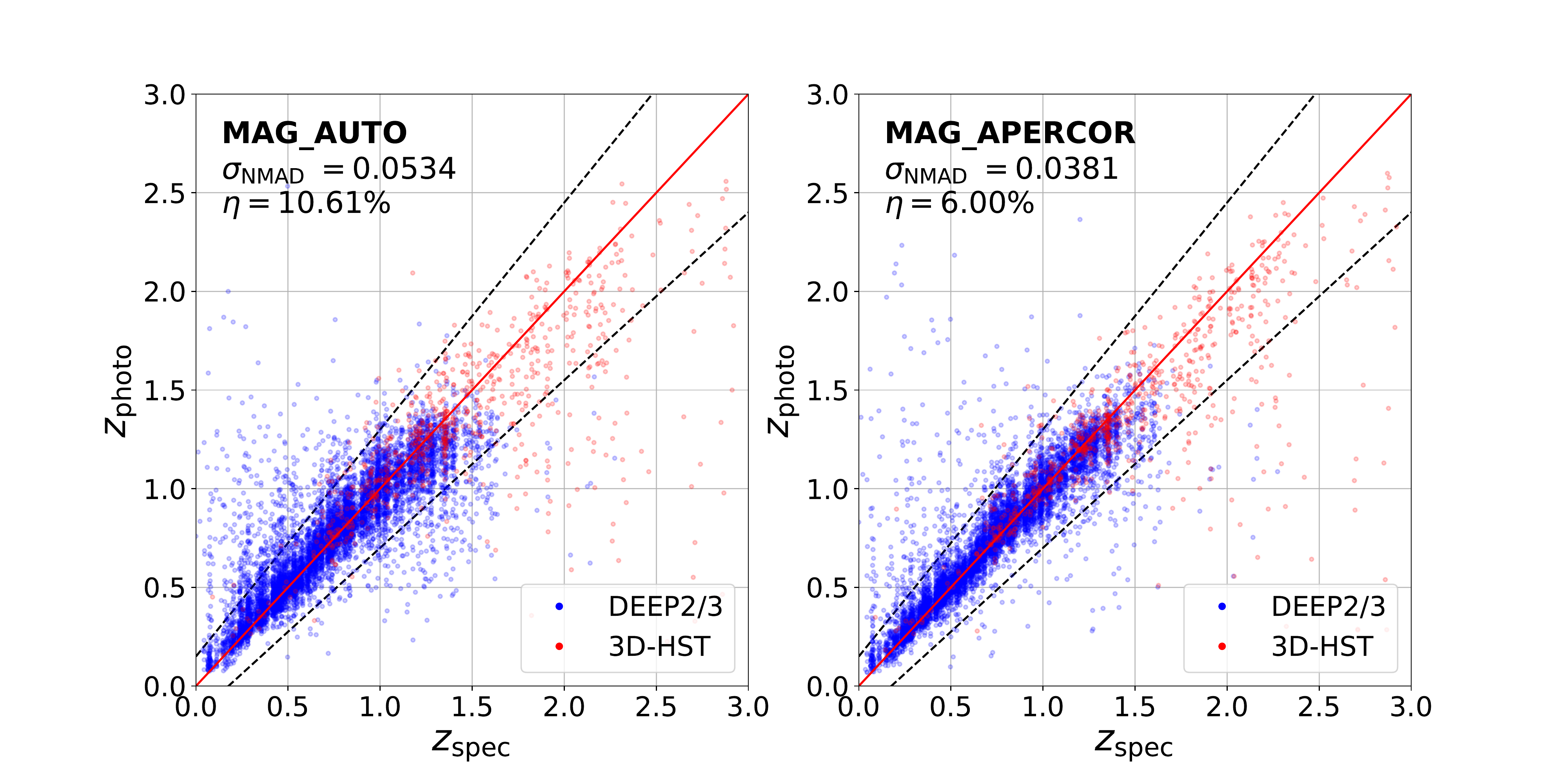}
    \caption{Photometric redshift vs spectroscopic or grism redshift using CFHTLS Wide field $ugriz$ and Subaru $Y$-band photometry. The red solid line corresponds to $z_{\mathrm{photo}}=z_{\mathrm{spec}}$. The dashed lines mark the boundary separating the outliers. The MAG\_APERCOR photometry produces photo-z's with significantly better accuracy than MAG\_AUTO.}
    \label{fig:photo-z_wide}
\end{figure*}

\begin{figure*}
    \centering
        \includegraphics[width=.8\textwidth]{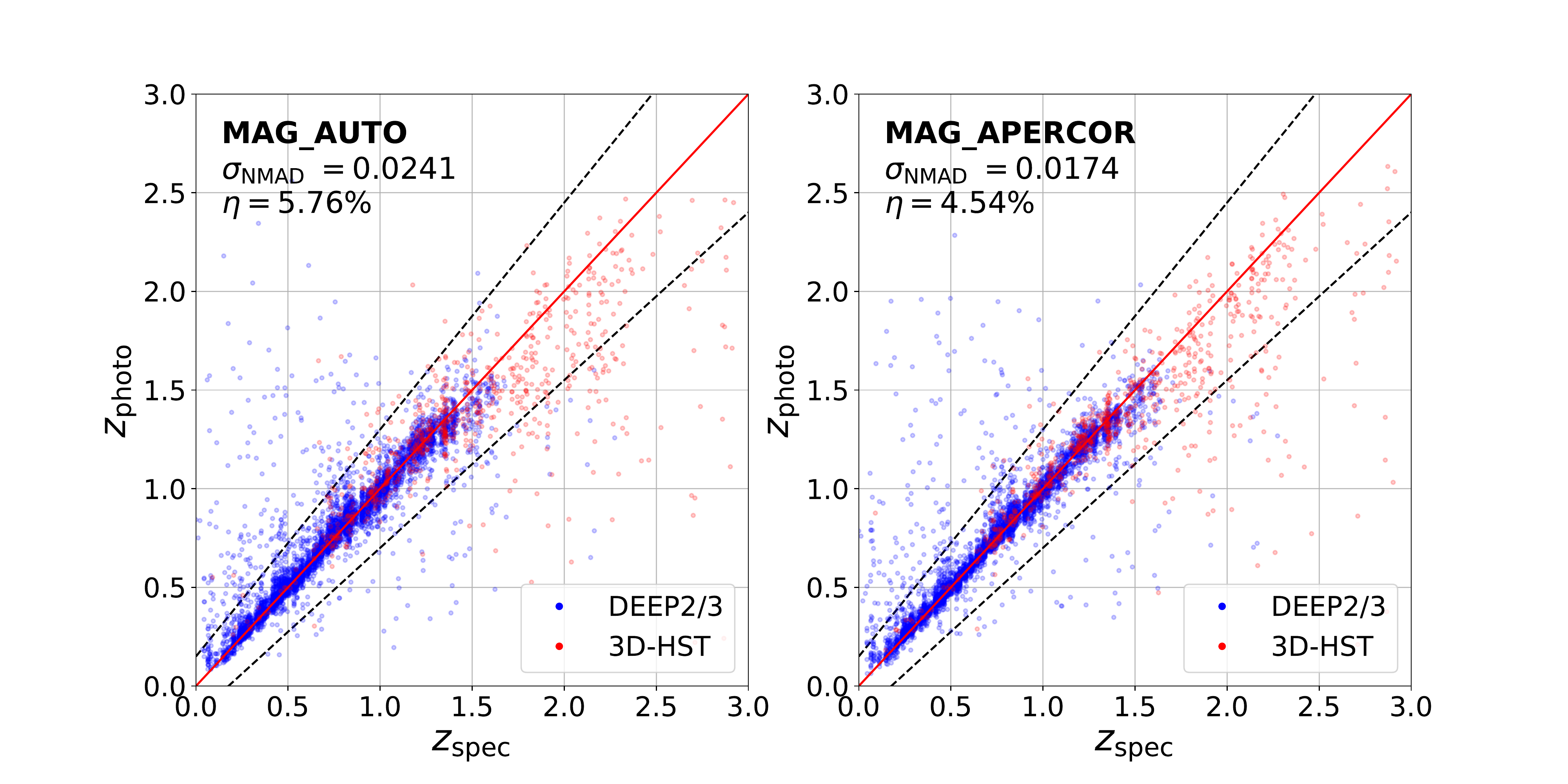}
    \caption{Same as Fig. \ref{fig:photo-z_wide}, but using CFHTLS Deep field photometry instead.}
    \label{fig:photo-z_deep}
\end{figure*}

Fig. \ref{fig:photo-z_wide} shows the photo-$z$ results using the CFHTLS Wide field photometry, and Fig. \ref{fig:photo-z_deep} shows the results with CFHTLS Deep field photometry. We find that using the MAG\_APERCOR photometry, we achieve photo-$z$ accuracy $\sigma_{\mathrm{NMAD}}=0.018$ and outlier fraction of 4.7\% in the CFHTLS Deep field, and $\sigma_{\mathrm{NMAD}}=0.039$ and 6.3\% outliers in the CFHTLS Wide field. This represents a significant improvement over MAG\_AUTO: $\sigma_{\mathrm{NMAD}}$ is reduced by 28\% in CFHTLS Wide and 27\% in CFHTLS Deep, and there is also a significant reduction in the outlier fraction.  The scatter in $\Delta z$ is larger at $z_{\mathrm{spec}}>1.4$ for both MAG\_AUTO and MAG\_APERCOR photometry and in both the Deep and Wide areas. This is due to both the small number of training objects in this redshift range, as well as the lack of available features (e.g., the $4000\text{\AA}$ break) in the optical.

As an additional validation of the MAG\_APERCOR photometry, we have performed similar photo-z tests using the CFHTLS photometry from the 3D-HST photometric catalogs (\citealt{Skelton2014}). In that work, the objects were detected with HST imaging, and forced photometry of these objects were performed on the CFHTLS Deep $ugriz$ images with an aperture of $1.2\arcsec$. We performed photo-z tests using the $ugriz$ photometry from \citet{Skelton2014} and redshifts from DEEP2/3 and 3D-HST, and for comparison we ran the same test using the CFHTLS Deep MAG\_AUTO and MAG\_APERCOR photometry in $ugriz$ bands for the same objects. We find that the \citet{Skelton2014} $ugriz$ photo-z's have very similar accuracy to the MAG\_APERCOR photo-z's, with the former having 2\% smaller $\sigma_\mathrm{NMAD}$ and 17\% fewer outliers. Both significantly outperform the MAG\_AUTO photo-z's, with the \citet{Skelton2014} $ugriz$ photo-z's having $37\%$ smaller $\sigma_\mathrm{NMAD}$ and $47\%$ fewer outliers than MAG\_AUTO.

The CFHTLS Deep field and the Subaru $Y$-band have depth similar to LSST 10-year data. Therefore this test also demonstrates that {in the magnitude and redshift range} of DEEP2/3, at least, it is possible for LSST to achieve the goal of $0.03(1+z)$ photo-$z$ accuracy as specified by the Science Requirements Document of the LSST Dark Energy Science Collaboration \citep{lsst_desc_srd}.

\section{Summary}
\label{sec:summary}
In this work we have presented a set of new catalogs with improved $ugrizY$ photometry and spectroscopic or grism redshifts in the Extended Groth Strip. We calibrated CFHTLS $ugriz$ photometry and Subaru $Y$-band photometry and also produced corrected aperture magnitudes. We combined the $ugrizY$ photometry with DEEP2/3 and 3D-HST redshifts. The $ugrizY$ photometry has depth similar to the LSST 10-year stack, and the catalogs will be useful for LSST photo-$z$ tests. All data is publicly available.

We have implemented a random forest photo-$z$ algorithm on our dataset, and found the photo-$z$ accuracy to be ${\sim} 2\%$ or better for the available spectroscopic sample in the deepest region, where the photometry has LSST-like depth. We also found significant improvement in photo-$z$ accuracy from the corrected aperture magnitude, indicating that our corrections provide a real improvement in the measurement of galaxy colors (as they tighten the color-redshift relation).

\section*{Acknowledgements}

We would like to thank Abhishek Prakash for his help with random forest photo-z estimation; Eric Gawiser for assistance with an early version of the corrected aperture photometry; and Dan Scolnic for providing useful information on photometric zero-point calibration. We also wish to thank the anonymous referee for their detailed reading and helpful comments on this paper.

CFHTLS is based on observations obtained with MegaPrime/MegaCam, a joint project of CFHT and CEA/IRFU, at the Canada-France-Hawaii Telescope (CFHT) which is operated by the National Research Council (NRC) of Canada, the Institut National des Science de l'Univers of the Centre National de la Recherche Scientifique (CNRS) of France, and the University of Hawaii. This work is based in part on data products produced at Terapix available at the Canadian Astronomy Data Centre as part of the Canada-France-Hawaii Telescope Legacy Survey, a collaborative project of NRC and CNRS.

This work is based in part on data collected at Subaru Telescope, which is operated by the National Astronomical Observatory of Japan.

Funding for the DEEP2 Galaxy Redshift Survey has been provided by NSF grants AST-95-09298, AST-0071048, AST-0507428,  AST-0507483,  and NASA LTSA grant NNG04GC89G.  DEEP3 was funded by NSF grants AST-08-08133, AST-08-07630, and AST-08-06732.  The development of the photometric redshift testbed presented in this paper was funded by DOE grant DE-SC0007914.

This work is partly based on observations taken by the 3D-HST Treasury Program (GO 12177 and 12328) with the NASA/ESA HST, which is operated by the Association of Universities for Research in Astronomy, Inc., under NASA contract NAS5-26555. 

Funding for the SDSS and SDSS-II has been provided by the Alfred P. Sloan Foundation, the Participating Institutions, the National Science Foundation, the U.S. Department of Energy, the National Aeronautics and Space Administration, the Japanese Monbukagakusho, the Max Planck Society, and the Higher Education Funding Council for England. The SDSS Web Site is http://www.sdss.org/.

The SDSS is managed by the Astrophysical Research Consortium for the Participating Institutions. The Participating Institutions are the American Museum of Natural History, Astrophysical Institute Potsdam, University of Basel, University of Cambridge, Case Western Reserve University, University of Chicago, Drexel University, Fermilab, the Institute for Advanced Study, the Japan Participation Group, Johns Hopkins University, the Joint Institute for Nuclear Astrophysics, the Kavli Institute for Particle Astrophysics and Cosmology, the Korean Scientist Group, the Chinese Academy of Sciences (LAMOST), Los Alamos National Laboratory, the Max-Planck-Institute for Astronomy (MPIA), the Max-Planck-Institute for Astrophysics (MPA), New Mexico State University, Ohio State University, University of Pittsburgh, University of Portsmouth, Princeton University, the United States Naval Observatory, and the University of Washington. 

The Pan-STARRS1 Surveys (PS1) and the PS1 public science archive have been made possible through contributions by the Institute for Astronomy, the University of Hawaii, the Pan-STARRS Project Office, the Max-Planck Society and its participating institutes, the Max Planck Institute for Astronomy, Heidelberg and the Max Planck Institute for Extraterrestrial Physics, Garching, The Johns Hopkins University, Durham University, the University of Edinburgh, the Queen's University Belfast, the Harvard-Smithsonian Center for Astrophysics, the Las Cumbres Observatory Global Telescope Network Incorporated, the National Central University of Taiwan, the Space Telescope Science Institute, the National Aeronautics and Space Administration under Grant No. NNX08AR22G issued through the Planetary Science Division of the NASA Science Mission Directorate, the National Science Foundation Grant No. AST-1238877, the University of Maryland, Eotvos Lorand University (ELTE), the Los Alamos National Laboratory, and the Gordon and Betty Moore Foundation.

DR acknowledges the support of the Science and Technology Facilities Council (STFC) through grant ST/P000541/1.


\clearpage

\appendix

\section{The DEEP3 Galaxy Redshift Survey}
\label{deep3}

The DEEP3 Galaxy Redshift Survey was a Large Multi-Annual Program allocated 25.5 nights of time on the DEIMOS spectrograph at the Keck 2 telescope to measure redshifts and properties of galaxies in the Extended Groth Strip. The combination of DEEP2 and DEEP3 provides roughly 18,000 redshifts in the portion of the Extended Groth Strip overlapping the greatest amount of multiwavelength data, including multiband imaging from \textit{HST} and \textit{Spitzer} and deep ACIS imaging with \textit{Chandra}.

DEEP3 includes observations of 56 DEIMOS slitmasks, tiling the central portion of the Extended Groth Strip and building upon the 120 slitmasks observed in the EGS as part of DEEP2. Observations for DEEP3 began in April 2008 and continued until May 2011. In total, DEEP3 targeted ${\sim} 7500$ sources, yielding ${\sim} 5000$ secure redshifts. Here, we provide details regarding the target selection, observations, and data reduction for DEEP3. The first public version of the DEEP3 redshift catalog as well as sky-subtracted one- and two-dimensional spectra of each target are available at \url{http://deep.ps.uci.edu/deep3/home.html}.

\begin{figure}
    \centering
        \includegraphics[width=\columnwidth]{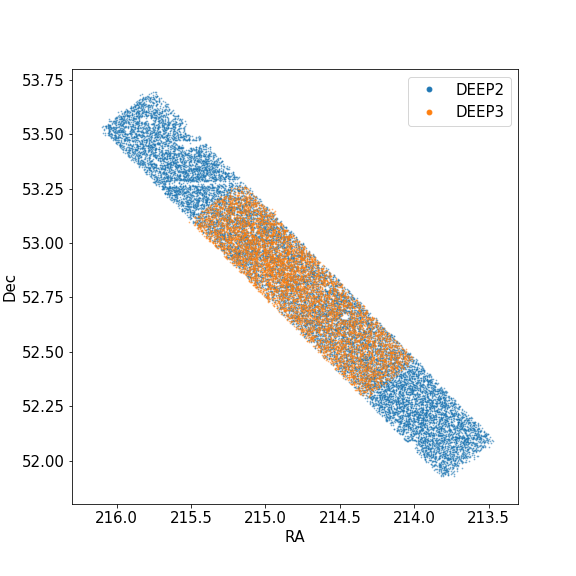}
    \caption{Sky coverage DEEP2 and DEEP3. The points are all the targets in the surveys, regardless of whether successful redshifts were obtained.}
    \label{fig:sky_coverage_deep2_deep3}
\end{figure}

\subsection{DEEP3 Target Samples}

The DEEP3 targeting strategy differs in a number of respects from the target selection strategy used by DEEP2 in the EGS, which was described by \citet{Newman13}.  First, at highest priority a set of objects were targeted based upon their unusual multiwavelength properties (e.g. X-ray or far-IR sources) in AEGIS imaging. A list of the various multi-wavelength sources observed and the bits used to identify objects from each sample in the DEEP3 redshift catalog can be found in Table {\ref{tab:deep3table}}. These objects were restricted to comprise only a small fraction of the overall sample to make sure that clustering measurements for the overall sample are not strongly affected.

Second, the ${\sim}35\%$ of $R_{AB} < 24.1$ galaxies which were unable to be targeted by DEEP2 due to slit collisions were assigned the next highest priority in maskmaking for DEEP3, providing in combination with DEEP2 a uniform sample of more than 90\% of all $R_{AB} < 24.1$ in the DEEP3 area that can be used for measurements of environment statistics and galaxy clustering. Third, at lowest priority  a `faint extension' of targets with $24.1 < r_{AB} < 25.5$ in CFHTLS imaging were targeted; the resulting sample was expected to be systematically incomplete but still yield a number of useful redshifts (in the end, roughly 40\% of $R>24.1$ targets in DEEP3 have secure redshift measurements).

A major difference between the strategies of DEEP2 and DEEP3 was the use of the 600-line grating on DEIMOS for the latter, instead of the 1200-line grating that was used in DEEP2.  The added spectral coverage to the blue from using a lower-resolution grating enables enhanced studies of line ratios, metallicities, AGN properties, K+A galaxy signatures, and Mg II wind absorption compared to DEEP2. Kinematic measurements of small-linewidth galaxies are not possible at this lower resolution, but these are already abundantly available in DEEP2. Tests prior to the start of DEEP3 found no reduction in redshift success using a lower resolution, despite the greater difficulty in resolving the [OII] doublet; in the end, DEEP3 obtained secure (ZQUALITY 3 or 4) for 69\% of galaxy targets with $R_{AB} < 24.1$, versus 73\% in DEEP2.  

The DEEP3 spectra cover a broader wavelength range than those from DEEP2, spanning 4550\AA\ -\ 9900\AA (with a central wavelength of 7200\AA). The GG455 order-blocking filter was used to limit flux blueward of 4550\AA, and each slitmask was observed for approximately one hour, depending on the observing conditions (i.e., transparency and/or seeing). Typical slitlengths were $\sim4$ -- $8\arcsec$, with a standard $1\arcsec$ slitwidth. A standard DEEP3 exposure consists of three 1200 s subexposures, which are used to remove cosmic rays and are then co-added to make a total exposure of 1 hour. 

\subsection{DEEP3 Maskmaking and Tiling Strategies}

The sky region covered by DEEP3 corresponds to the central 50\% (in the long direction) of the DEEP2 region of EGS, as shown in Fig. \ref{fig:sky_coverage_deep2_deep3}. This region corresponds to the intersection of the most important multi-wavelength surveys in the field, including coverage with {\it Spitzer} IRAC and MIPS, {\it HST}/ACS, {\it Chandra}, and {\it GALEX}. VLA 20cm data are poorer in the lower part of the strip owing to interference by the bright source 3C295, providing additional reason to avoid the southern end of DEEP2 for this project.

DEEP3 masks were spaced 1.5 arcmin apart, rather than 1 arcmin as in DEEP2, to match the density of targets for the program (since the majority of $R_{AB} < 24.1$ galaxies were already targeted by DEEP2). DEEP3 masks cover a strip that is $15\arcmin$ wide with DEIMOS, as DEEP2 did, even though the region covered by {\it Spitzer} and {\it HST} is only $10\arcmin$ wide. This is needed in order to create an overhang region that extends at least $2.5\arcmin$ beyond the prime imaging area in all directions. This buffer zone allows us to measure environmental densities for all objects in the prime zone free of edge effects. Without it, only half of the $10\arcmin$ wide zone would be suitable for environmental studies.  The strip covered by DEEP3 masks is 1 degree long, versus 2 degrees for DEEP2.

Targets were placed on masks using a modified version of the maskmaking algorithms described in \citet{Newman13}.  For multiwavelength-selected objects, a wide variety of selection algorithms were used (cf, Table \ref{tab:deep3table}).  The priority of objects from this table was used as the selection weight ($W$ as defined in \citet{Newman13}) for them.  

For $R_{AB} < 24.1$ objects, the selection is similar but not identical to that used for DEEP2 in the Groth Strip.  As before, objects were required to meet the magnitude limits of the DEEP2 survey ($18.25 < R_{AB} < 24.1$; to have at least 20\% probability of being a galaxy ($p_{gal}>0.2$, as defined in \citet{Newman13}); and to have no imaging pixel flags set in the $R$ band.  Unlike in DEEP2, however, objects on either side of the DEEP2 color selection cuts were treated identically for DEEP3, and objects with nondetections in the $B$ or $I$ band or with low surface brightnesses were included in the sample.  DEEP2-like objects received a magnitude-based target selection weight $W_R$ (again, as defined in \citet{Newman13}) given by $\tt{min}(0.75 \times 10^{-0.4*(R-24.1)},1)$; this function falls from 1 at $R \leq 23.8$ to 0.75 at $R=24.1$; this is the same functional form used for higher-redshift objects in the EGS in DEEP2. Objects which were previously observed by DEEP2 but received non-secure redshifts in visual inspection ($Q=2$) were included in the sample for DEEP3, but with $W_R$ lowered by a factor of two (so the maximum possible $W_R$ for such objects was 0.5, instead of 1 for an unobserved DEEP2-like object).  Unlike in DEEP2, $W_R$ was not multiplied by the galaxy probability from star-galaxy separation, so the overall selection priority is $W = W_R$ for this sample.

For the ``faint extension'' of $R_{AB} > 24.1$ objects, the selection procedure was modified since CFHTLS data was used. Specifically, the CFHT Sextractor MAG\_AUTO $r$ magnitudes from the 2008A Megapipe CFHTLS catalogs produced by Stephen Gwyn (\url{http://www.cadc-ccda.hia-iha.nrc-cnrc.gc.ca/en/megapipe/cfhtls/index.html}; \citealt{Gwyn08}) were used to select targets for DEEP3. Eligible faint extension targets had $R_{AB} > 24.1$ in the DEEP2 photometric catalogs and $r > 23.5$ in CFHTLS, or a nondetection in the DEEP2 catalogs and $r > 24.22$ in CFHTLS (reflecting the average offset between DEEP2 $R$ and CFHTLS $r$), $r < 25.62$ in CFHTLS, and no $r$-band pixel flags set in the CFHTLS imaging.  These objects were given a weight $W_R = 0.2 \times \tt{min}(0.25 \times 10^{-0.2*(r-25.5)},1)$; this function falls from 0.090 at $r=24.22$ to 0.047 at $r=25.62$.  

Maskmaking then proceeded via the same procedure used by DEEP2, with the central region (in the wavelength direction) of each mask populated in a first mask, and outer portions populated second; there were only two minor differences in the procedure.  First, for DEEP3, a minimum slit length of 4 arcseconds, rather than 3 as in DEEP2, was used.

Second, in cases where multiple objects conflicted with each other such that they could not all be observed simultaneously, the target to be observed is chosen randomly.  For DEEP2, this was done by generating a random value between 0 and 1 for each object and choosing the one with highest random value.  For DEEP3, this behavior was altered to ensure selection of high-priority targets.  Specifically, for objects with weight $W > 1$, the object weight is multiplied by a random number uniformly distributed between 0.75 and 1; for objects with weights between 0.25 and 1, the object is assigned a random number uniformly distributed between 0 and 1 with no multiplication by weight; and for objects with weights below 0.25, the object weight is multiplied by a random number between 0 and 1.  Apart from these minor differences, maskmaking proceeded as in DEEP2.

\subsection{Data Reduction and Catalogs}

The DEIMOS data were reduced using a version of the DEEP2 DEIMOS {\tt spec2d} pipeline slightly modified to improve handling of 600-line grating data, yielding sky-subtracted 1-d and 2-d spectra for each object. Redshifts were then measured using the DEEP2 {\tt spec1d} Redshift Pipeline, with each redshift inspected by eye by at least one individual and assigned a quality code. The quality code system used is the same as DEEP2. ZQUALITY = -2 indicates a spectrum with data so poor for instrumental reasons that it was effectively not observed. ZQUALITY = -1 is used for stars.  ZQUALITY = 1 indicates a spectrum with such poor signal-to-noise that it is unlikely a redshift could be recovered, and ZQUALITY = 2 indicates that a reliable redshift could not be established for reasons specified in the COMMENT field.  Finally, ZQUALITY = 3 indicates a secure redshift ($>95\%$ probability of being correct), and ZQUALITY = 4 indicates highly secure cases ($>99\%$ probability of being correct). More details on the DEEP2 code used and the basic properties included in redshift catalogs may be found in \citet{Newman13}.

The DEEP3 redshift catalog adds a new tag (or column) for each object, EGSFLAGS, which has no analog in the DEEP2 redshift catalog. This tag provides information about which objects belong to which input target list. Unlike DEEP2, which employed a single set of selection cuts on $R_{AB} < 24.1$ galaxies, DEEP3 has targeted a variety of sources pulled from a variety of input catalogs provided by collaborators. Table \ref{tab:deep3table} shows the breakdown of the target list according to the flag values (and associated target lists). Many objects will have been eligible for targeting based on multiple reasons; e.g., a source might be both a ``FIDEL 24 $\micron$ priority 1'' source and a ``DEEP2 previously untargeted'' object. In such cases, all of the relevant flags are set --- for example, a Chandra source which is also a power-law AGN candidate will have both the $2^4$ and $2^{10}$ bits set, corresponding to an EGSFLAGS value of 1040. In other words, the EGSFLAGS value is an integer value containing the bitwise OR of all of the flag values pertaining to a given object. As can be seen from the table, the fractions of objects selected varied from survey to survey both due to varying target priorities (as listed in the table) and varying sky coverage; for instance, many Chandra sources fell at the ends of the slitmasks and thus were not able to be assigned a slit.


\begin{landscape}
\begin{table}
\caption{Table of all target samples included in DEEP2.  For each sample, we specify the corresponding bit in EGSFLAGS; the nature of the sample; the individual who provided it; the priority assigned in target selection (as input to the target selection procedure described in \citet{Newman13}; the number of targets within the overall DEEP3 footprint; the number of targets whose spectra were obtained; and the fraction of objects in the catalog which were targeted.}
\label{tab:deep3table}
\begin{tabular}{lllllllll}
Bit & Target Class & Contact & Median & \# in& \# with & \# with  & Fraction  & Fraction of targets \\[-2mm]
 & & &  priority & mask area & spectrum & secure z & targeted & with secure z \\[4mm]
$2^{ 0}$ & Spitzer/MIPS 70 $\micron$ sources, priority 1 & Mark Dickinson & 4$\times 10^{14}$ & 300 & 204 & 144 & 0.680 & 0.706 \\
$2^{ 1}$ & Spitzer/MIPS 70 $\micron$ sources, priority 2 & Mark Dickinson & 2$\times 10^{14}$ & 83 & 46 & 32 & 0.554 & 0.696 \\
$2^{ 2}$ & Spitzer/MIPS 24 $\micron$ sources, priority 1 & Mark Dickinson & 4$\times 10^{14}$ & 86 & 47 & 27 & 0.547 & 0.574 \\
$2^{ 3}$ & Spitzer/MIPS 24 $\micron$ sources, priority 2 & Mark Dickinson & 2$\times 10^{14}$ & 20 & 15 & 9 & 0.750 & 0.600 \\
$2^{ 4}$ & Chandra sources & Kirpal Nandra & 8$\times 10^8$ & 205 & 141 & 72 & 0.688 & 0.511 \\
$2^{ 5}$ & z\textless{}2 Massive Galaxies from AEGIS \citep{Conselice2007} & Christopher Conselice & 5$\times 10^{11}$ & 441 & 255 & 192 & 0.578 & 0.753 \\
$2^{ 6}$ & VLA 20 cm sources \citep{willner2012} & Robert Ivison & 8$\times 10^{14}$ & 125 & 93 & 49 & 0.744 & 0.527 \\
$2^{ 7}$ & Bright Akari/IRC 15 $\micron$ sources & Myungshin Im & 1$\times 10^{14}$ & 65 & 55 & 39 & 0.846 & 0.709 \\
$2^{ 8}$ & Faint Akari/IRC 15 $\micron$ sources & Myungshin Im & 5$\times 10^{5}$ & 91 & 44 & 16 & 0.484 & 0.364 \\
$2^{ 9}$ & Spitzer IRS targets & Jiasheng Huang & 2$\times 10^{14}$ & 4 & 3 & 0 & 0.750 & 0.000 \\
$2^{10}$ & Spitzer/IRAC Power-law AGN candidates & Jiasheng Huang & 2.5$\times 10^8$ & 76 & 49 & 18 & 0.645 & 0.367 \\
$2^{11}$ & VLA 6 cm sources \citep{willner2006} & Steven Willner & 4$\times 10^{14}$ & 90 & 73 & 38 & 0.811 & 0.521 \\
$2^{12}$ & Spitzer/IRAC-identified AGN & David Rosario & 1$\times 10^{14}$ & 24 & 22 & 9 & 0.917 & 0.409 \\
$2^{13}$ & New strong lens systems & Leonidas Moustakas & 2$\times 10^{15}$ & 3 & 2 & 2 & 0.667 & 1.000 \\
$2^{14}$ & Spitzer IRS object & Christopher Willmer & Not in area & N/A & N/A & N/A & N/A & N/A \\
$2^{15}$ & Dual AGN Candidates & Brian Gerke & 2$\times 10^{15}$ & 7 & 7 & 7 & 1.000 & 1.000 \\
$2^{16}$ & DEEP2 objects - previously untargeted & Jeffrey Newman & 8.2$\times 10^{11}$ & 7454 & 4420 & 3181 & 0.593 & 0.720 \\
$2^{17}$ & DEEP2 objects - previously targeted & Jeffrey Newman & 0.5 & 2595 & 1205 & 605 & 0.464 & 0.502 \\
$2^{18}$ & DEEP3 faint extension & Jeffrey Newman & 0.0595 & 30868 & 1346 & 539 & 0.044 & 0.400 \\
$2^{19}$ & DEEP2 strong lens reobservations & Jeffrey Newman & 1$\times 10^{14}$ & 12 & 12 & 12 & 1.000 & 1.000 \\
$2^{20}$ & SNLS supernova hosts - high priority & Saul Perlmutter & 2000 & 3 & 1 & 1 & 0.333 & 1.000 \\
$2^{21}$ & SNLS supernova hosts - low priority & Saul Perlmutter & 1$\times 10^{14}$ & 42 & 29 & 23 & 0.690 & 0.793 \\
$2^{22}$ & AEGIS-X sources & Kirpal Nandra & 8$\times 10^{14}$ & 141 & 96 & 63 & 0.681 & 0.656
\end{tabular}
\end{table}
\end{landscape}

\section{$Y$-band SExtractor parameters}
\label{sec:SExtractor}
Source catalogs in the $Y$-band were obtained by running SExtractor on the $Y$-band images. The SExtractor parameters used for the deep pointing are listed in section \ref{sec:sexparameters}. For the shallow pointing, only a few parameters were altered; these are listed at the end of the table. The ``PHOT\_APERTURES'' parameters specify the aperture diameters of the MAG\_APER photometry, which we use to compute the MAG\_APERCOR photometry. Note that SExtractor (version 2.19.5) cannot produce more than 30 aperture magnitudes, so we had to separate the apertures into two parameter files (but with the same maximum aperture size to ensure the same set of detections) and run them separately.

\subsection{SExtractor parameters}
\label{sec:sexparameters}
\begin{singlespace}

\scriptsize
\begin{verbatim}
 
SExtractor parameters for the deep pointing
 
#------------------------------- Extraction ----------------------------------
 
DETECT_TYPE      CCD            # CCD (linear) or PHOTO (with gamma correction)
DETECT_MINAREA   3              # min. # of pixels above threshold
DETECT_MAXAREA   6400

DETECT_THRESH    2.0            # <sigmas> or <threshold>,<ZP> in mag.arcsec-2
ANALYSIS_THRESH  2.0            # <sigmas> or <threshold>,<ZP> in mag.arcsec-2
THRESH_TYPE      RELATIVE
 
FILTER           Y              # apply filter for detection (Y or N)?
FILTER_NAME      gauss_2.5_5x5.conv   # name of the file containing the filter
 
DEBLEND_NTHRESH  64             # Number of deblending sub-thresholds
DEBLEND_MINCONT  0.001          # Minimum contrast parameter for deblending
 
CLEAN            Y              # Clean spurious detections? (Y or N)?
CLEAN_PARAM      1.0            # Cleaning efficiency
 
#-------------------------------- WEIGHTing ----------------------------------

WEIGHT_GAIN      N              # If true, weight maps are considered as gain maps.
WEIGHT_TYPE      MAP_RMS        # type of WEIGHTing: NONE, BACKGROUND,
                                # MAP_RMS, MAP_VAR or MAP_WEIGHT
WEIGHT_IMAGE     weight_maps/BACKGROUND_RMS_SIZE_16.FITS    # weight-map filename

#-------------------------------- FLAGging -----------------------------------

FLAG_IMAGE       edge_flag.fits # filename for an input FLAG-image
FLAG_TYPE        MOST           # flag pixel combination: OR, AND, MIN, MAX
                                # or MOST

#------------------------------ Photometry -----------------------------------

PHOT_APERTURES   9, 10, 11, 12, 13, 14, 15, 16, 17, 18, 19, 20, 21, 22, 23, 24, 
25, 26, 27, 28, 29, 30, 31, 32, 33, 34, 35, 36, 37, 38, 39, 40, 41, 42, 43, 44, 
45, 46, 47, 48, 49, 50, 51, 52, 53, 54, 55, 56
                                # MAG_APER aperture diameter(s) in pixels
PHOT_AUTOPARAMS  2.5, 3.5       # MAG_AUTO parameters: <Kron_fact>,<min_radius>
PHOT_PETROPARAMS 2.0, 3.5       # MAG_PETRO parameters: <Petrosian_fact>,
                                # <min_radius>
PHOT_AUTOAPERS   20.0,20.0      # <estimation>,<measurement> minimum apertures
                                # for MAG_AUTO and MAG_PETRO
PHOT_FLUXFRAC    0.2, 0.5, 0.8  #Fraction of FLUX AUTO defining each element of 
the FLUX RADIUS vector.

SATUR_LEVEL      36000.0        # level (in ADUs) at which arises saturation
SATUR_KEY        SATURATE       # keyword for saturation level (in ADUs)
 
MAG_ZEROPOINT    31.2           # magnitude zero-point
MAG_GAMMA        4.0            # gamma of emulsion (for photographic scans)
GAIN             1              # detector gain in e-/ADU
GAIN_KEY         GAIN           # keyword for detector gain in e-/ADU
PIXEL_SCALE      0              # size of pixel in arcsec (0=use FITS WCS info)
 
#------------------------- Star/Galaxy Separation ----------------------------
 
SEEING_FWHM      0.648          # stellar FWHM in arcsec
STARNNW_NAME     default.nnw    # Neural-Network_Weight table filename
 
#------------------------------ Background -----------------------------------
 
BACK_TYPE        AUTO           # AUTO or MANUAL
BACK_VALUE       0.0            # Default background value in MANUAL mode
BACK_SIZE        128            # Background mesh: <size> or <width>,<height>
BACK_FILTERSIZE  5              # Background filter: <size> or <width>,<height>
BACKPHOTO_TYPE   LOCAL
BACKPHOTO_THICK  24

#--------------------- Memory (change with caution!) -------------------------
 
MEMORY_OBJSTACK  3000           # number of objects in stack
MEMORY_PIXSTACK  9000000        # number of pixels in stack
MEMORY_BUFSIZE   1024           # number of lines in buffer


\end{verbatim}

The following parameters are for the shallow pointing: 

\begin{verbatim}
DETECT_MINAREA   5
SATUR_LEVEL      280000.0
SEEING_FWHM      0.625
\end{verbatim}

\end{singlespace}

\section{Aperture correction procedures}
\label{sec:app_MAG_APERCOR}

Two assumptions are made in determining our aperture corrections. The first is that all objects have a circular symmetry and their light profiles can be described by a Moffat profile (described in more detail below). The second assumption is that in each band in each pointing, the parameters describing the Moffat profile only depend on the half-light radius and that they are smooth functions of this quantity. Under these assumptions, we can measure the flux in a small aperture and use the Moffat profile appropriate for a given object's half-light radius to extrapolate the total flux. We perform aperture corrections separately for each band in each pointing so that we can account for differences between seeing in each image.

The Moffat light profile is described by the equation
\begin{equation}
\label{eq:moffat}
    I(r; \alpha, \beta) = \frac{\beta-1}{\pi \alpha^2}\left[1+\left(\frac{r}{\alpha}\right)^2\right]^{-\beta},
\end{equation}
where $I$ denotes the flux density and $r$ is the angular distance from the center of the source. There are two free parameters: $\alpha$ determines the width of the profile and $\beta$ determines its shape. If $\beta$ is small, the light profile includes more flux at larger radii (larger ``wings''), while $\beta \to \infty$ corresponds to a Gaussian profile. In this formula, the light profile is normalized so that the total flux is 1. The fraction of flux inside radius $r$ is then
\begin{equation}
\label{eq:fraction}
    \mathrm{frac}(r) = \int_0^r 2\pi x I(x) dx = 1 - \alpha^{2(\beta-1)}\left(\alpha^2 + x^2\right)^{1-\beta}.
\end{equation}

A measurement of the half-light radius from SExtractor is provided by CFHTLS. In principle, we can determine $\alpha$ by solving equation \ref{eq:fraction} for the case $I(R_{1/2}; \alpha, \beta) = 1/2$, where $R_{1/2}$ is the half-light radius, leaving only one free parameter, $\beta$. However, we found that the ``half-light'' radius measured by SExtractor does not capture exactly half of the total flux, so we treat $\alpha$ as a free parameter as well. In the rest of this section we use $R_{1/2}$ and the word radius to refer to the SExtractor-measured half-light radius rather than the value derived from the Moffat fit.

One set of $\alpha$ and $\beta$ is enough to characterize the light profiles of stars since they have essentially the same light profile (i.e. the PSF). Galaxies have different light profiles, so we divide galaxies into radius bins and find the optimal $\alpha$ and $\beta$ for each bin. The bin sizes are $0.0558\arcsec$ for $u$ and $z$ bands, $0.0372\arcsec$ for $g, r, i$ bands and $0.03\arcsec$ for $Y$-band. The smallest bin is set by the PSF (stars) and the largest bin has a radius of $1.1\arcsec$--$1.2\arcsec$. We use the CFHTLS ``flag'' column for star-galaxy separation.

To avoid large radius errors in bands with low S/N, and also to reduce errors in colors (e.g. $u-g$) by ensuring consistent treatment of radii, we binned objects according to their $r$-band radii when determining the aperture correction for each CFHTLS passband. For the $Y$-band aperture correction the $Y$-band radius was used for binning as many objects are not detected in $r$.
For each radius bin, we compute the average curve of growth of flux as a function of radius by simply averaging the curve of growth of the individual objects within that bin.

 CFHTLS provides SExtractor aperture magnitudes (MAG\_APER) for aperture radii ranging from 5 pixels to 30 pixels in 1 pixel spacing; we use these magnitudes for the curve of growth calculations. For the $Y$-band we also produced similar SExtractor aperture magnitudes; see Appendix \ref{sec:SExtractor} for details of the $Y$-band aperture magnitudes.

\begin{figure*}
    \centering
    \begin{subfigure}[b]{0.33\textwidth}
        \includegraphics[width=\textwidth]{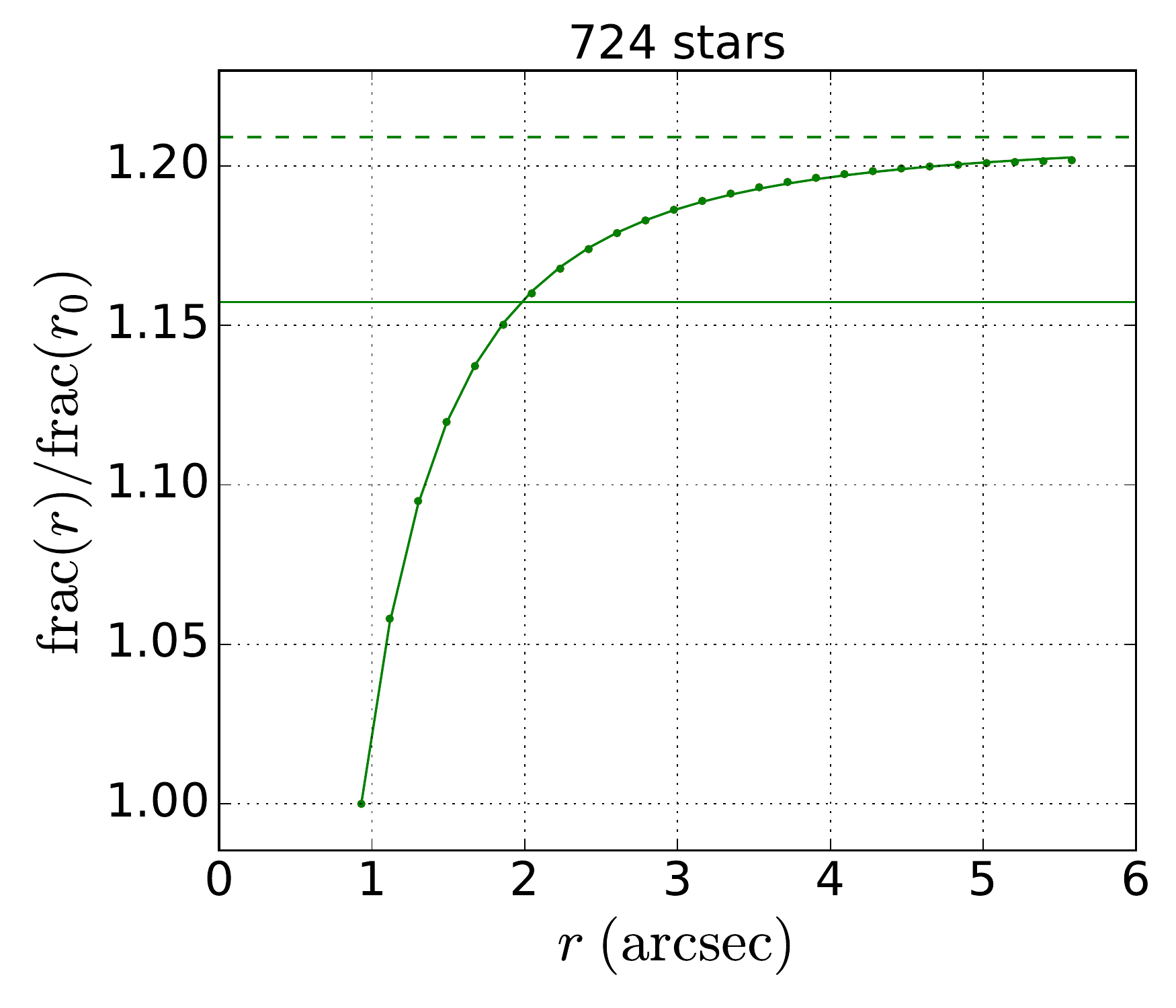}
        \caption{}
    \end{subfigure}
    \begin{subfigure}[b]{0.33\textwidth}
        \includegraphics[width=\textwidth]{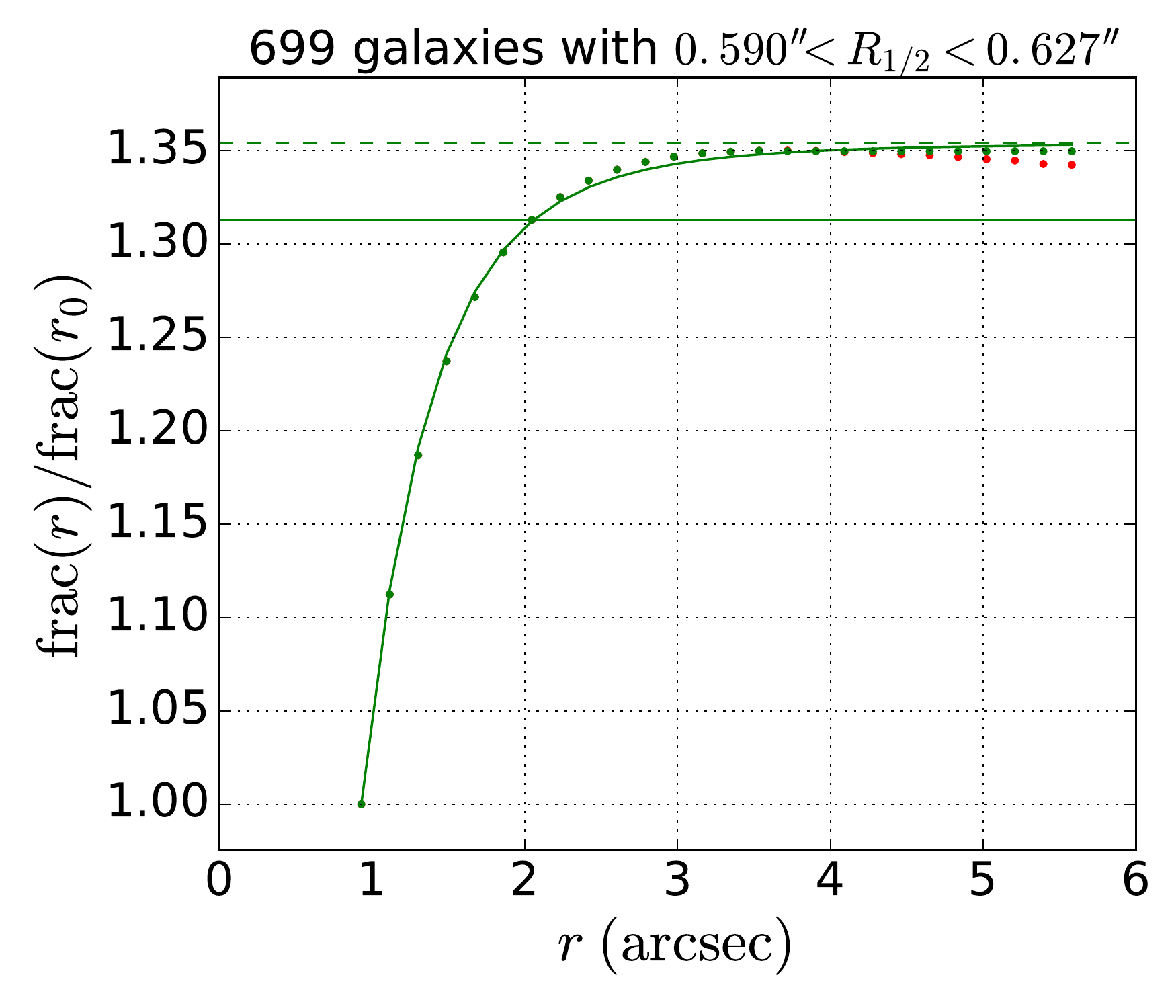}
        \caption{}
    \end{subfigure}
    \begin{subfigure}[b]{0.33\textwidth}
        \includegraphics[width=\textwidth]{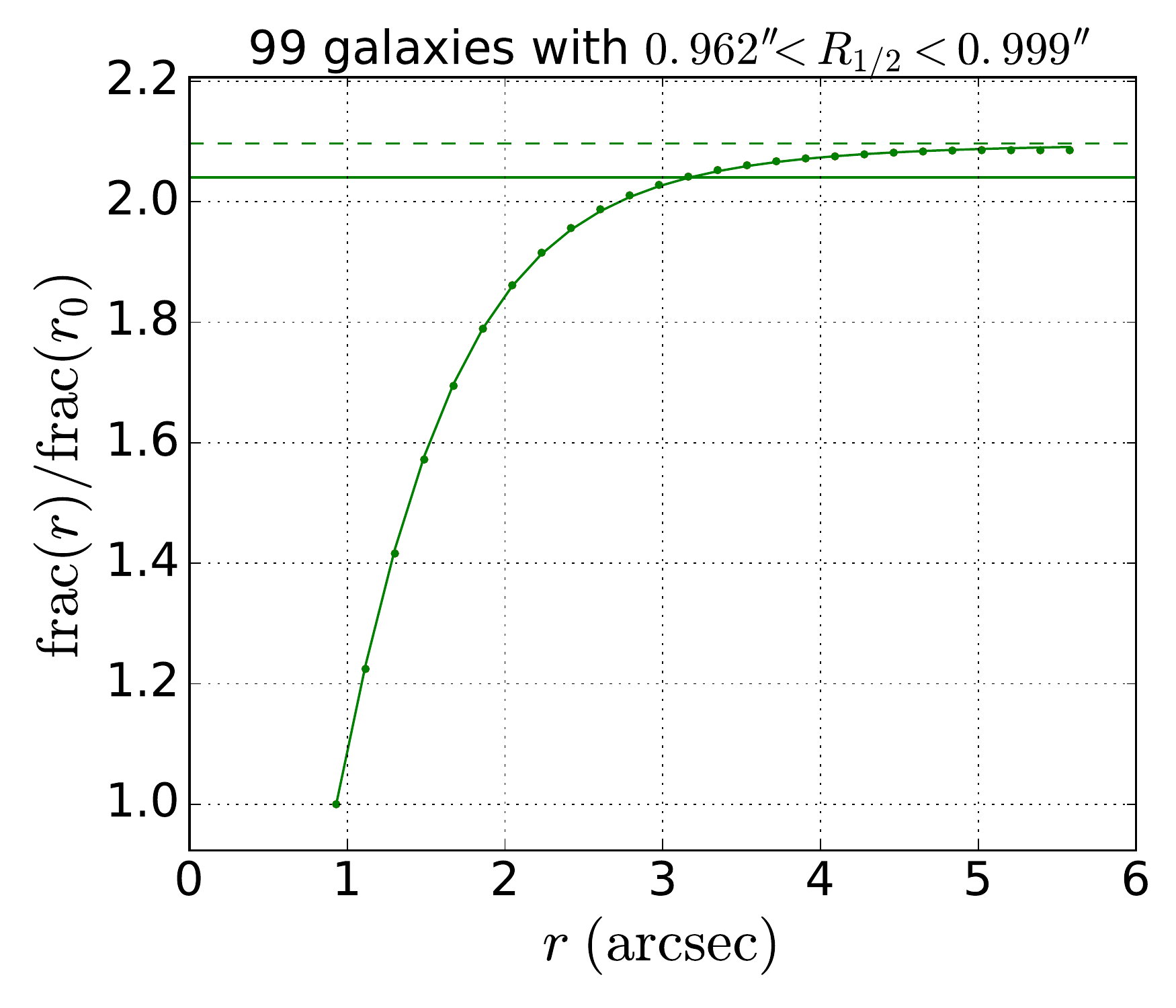}
        \caption{}
    \end{subfigure}\\
    \begin{subfigure}[b]{0.33\textwidth}
        \includegraphics[width=\textwidth]{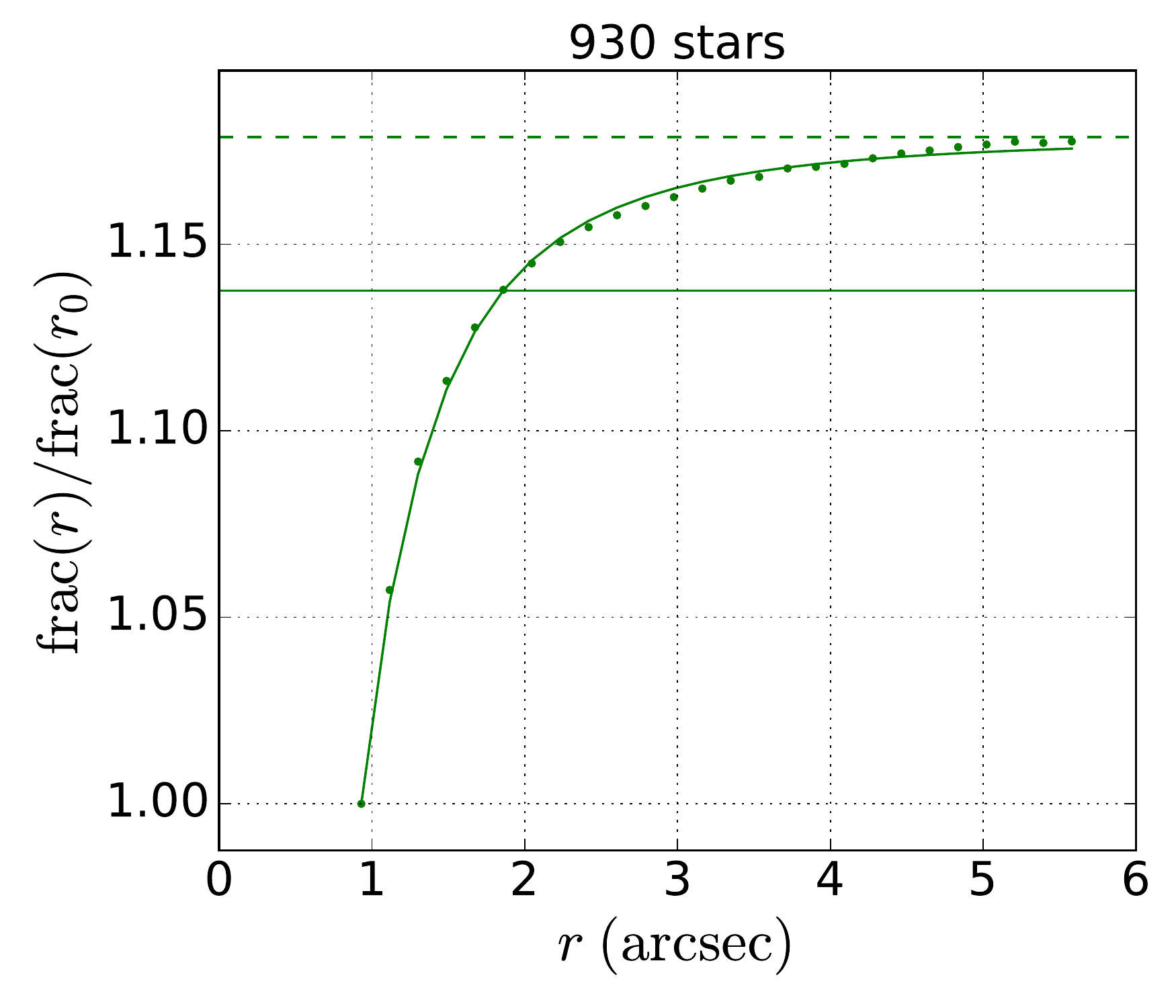}
        \caption{}
    \end{subfigure}
    \begin{subfigure}[b]{0.33\textwidth}
        \includegraphics[width=\textwidth]{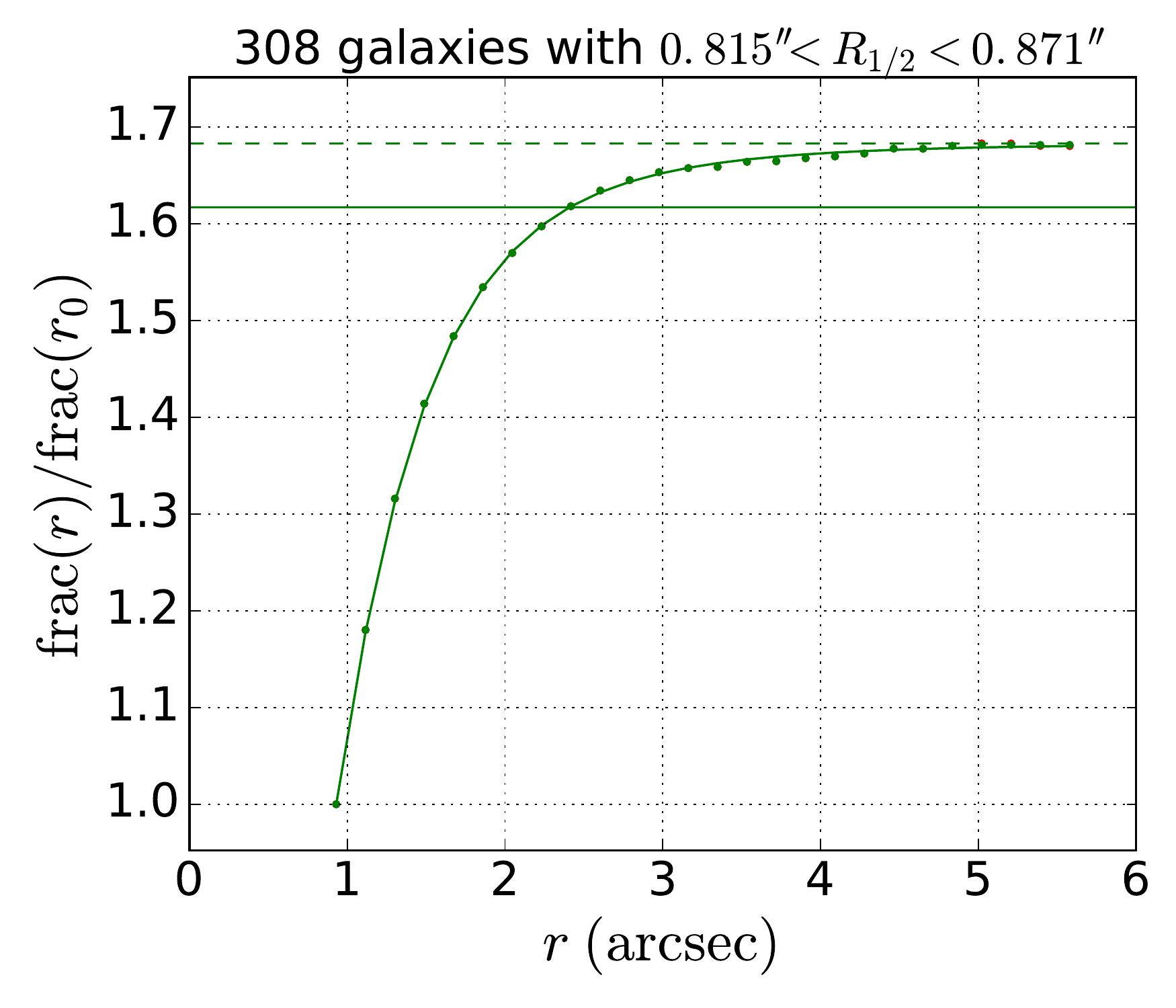}
        \caption{}
    \end{subfigure}
    \begin{subfigure}[b]{0.33\textwidth}
        \includegraphics[width=\textwidth]{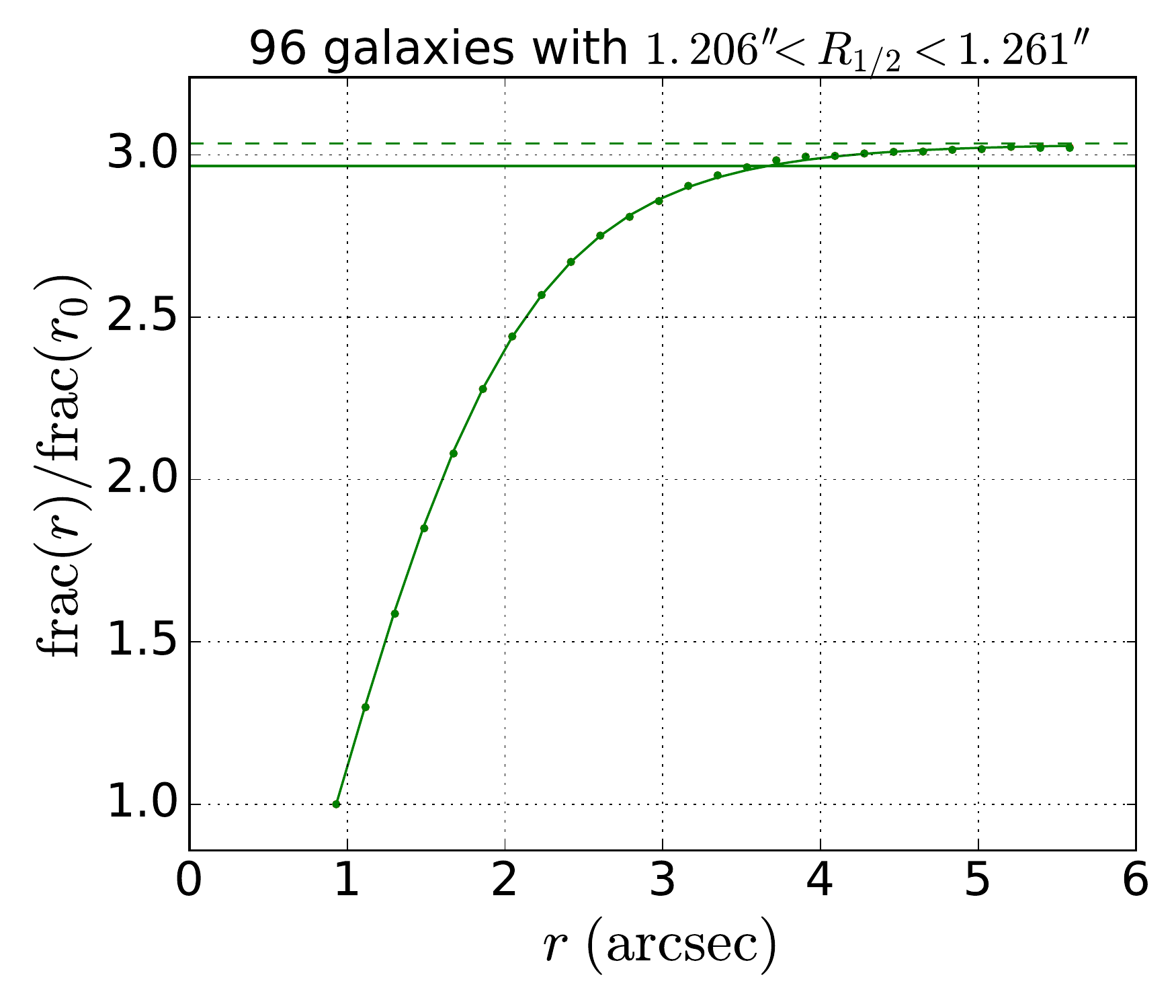}
        \caption{}
    \end{subfigure}\\
    \begin{subfigure}[b]{0.33\textwidth}
        \includegraphics[width=\textwidth]{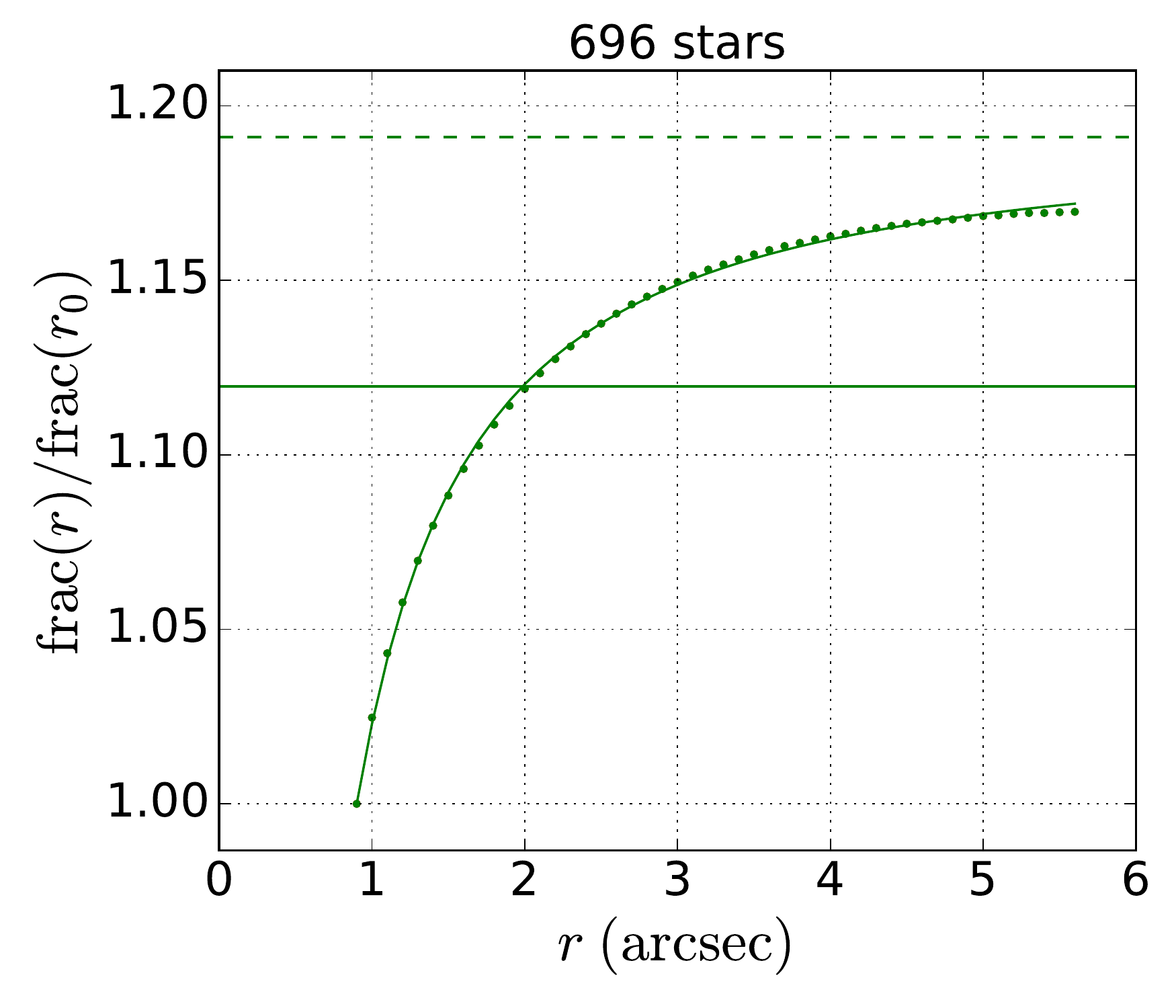}
        \caption{}
    \end{subfigure}
    \begin{subfigure}[b]{0.33\textwidth}
        \includegraphics[width=\textwidth]{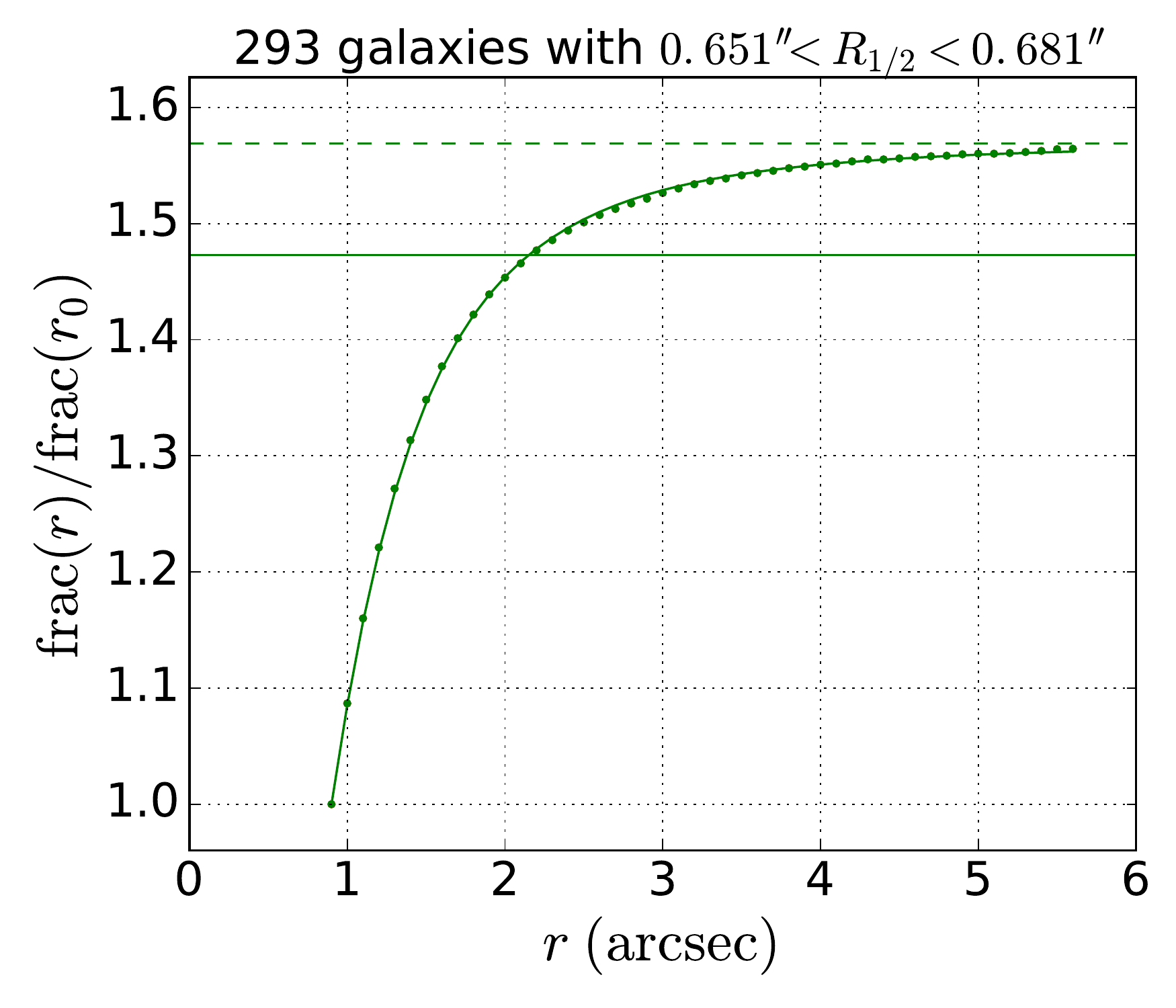}
        \caption{}
    \end{subfigure}
    \begin{subfigure}[b]{0.33\textwidth}
        \includegraphics[width=\textwidth]{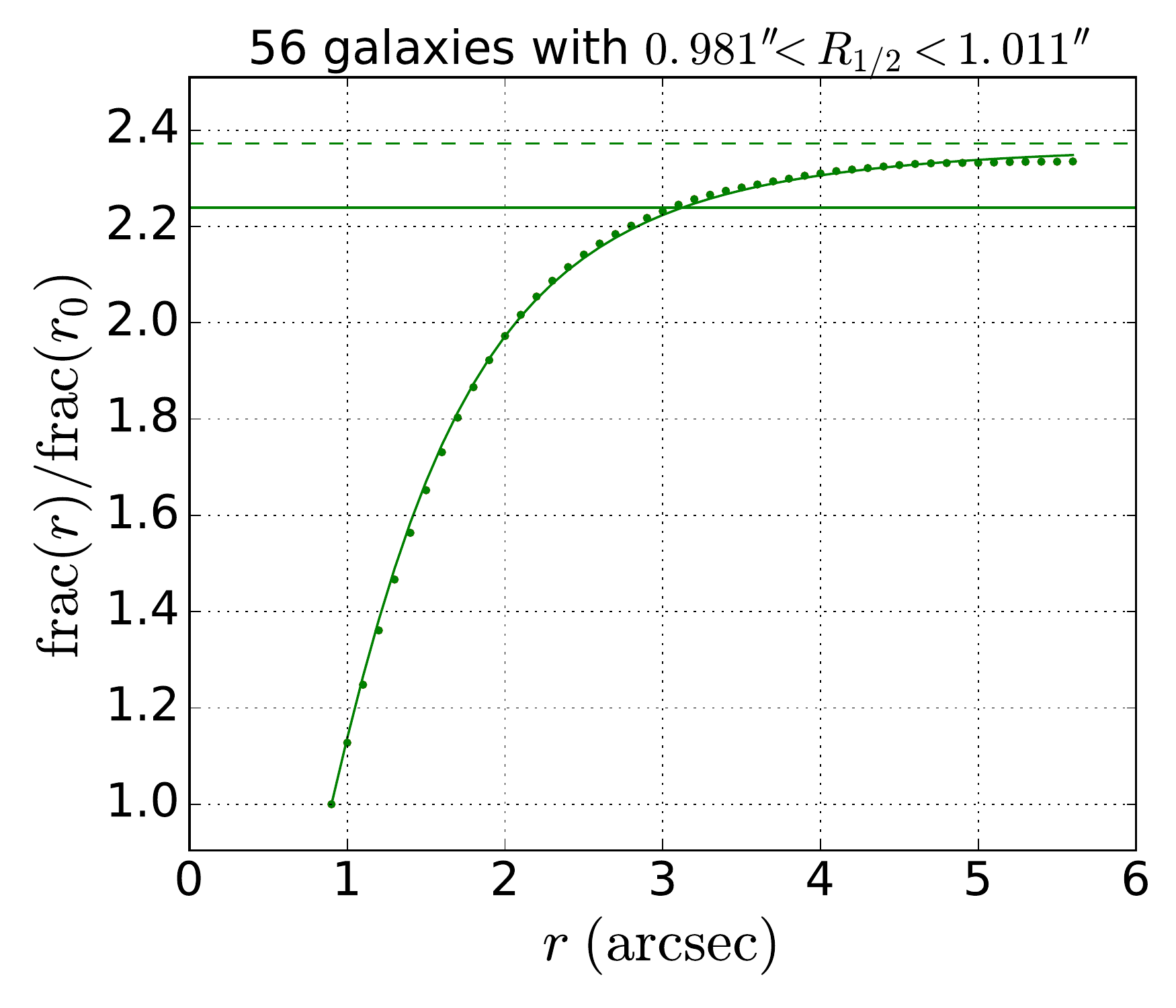}
        \caption{}
    \end{subfigure}
    \caption{Examples of curve of growth and its Moffat fit. The $Y$-axis is the ratio of the flux in aperture radius $r$ to the flux in the fixed aperture radius $r_0$. Points are the observed flux ratio for each radius bin. The solid curve is the Moffat fit. The solid horizontal line is the ratio of the flux in MAG\_AUTO to the flux in the fixed aperture radius $r_0$, and the dashed horizontal line is the predicted flux ratio for an infinitely large aperture. (a), (b) and (c) show CFHTLS D3 $i$-band. (d), (e) and (f) show CFHTLS W3-0-2 $u$-band. (g), (h) and (i) show the Subaru $Y$-band deep pointing. In (b), the flux ratio decreases at large apertures (red points) due to non-zero background, and it is corrected by extrapolating using the maximum flux ratio (green points). Such non-zero background might carry a different sign, showing as large increase of flux ratio at large apertures, although in this case it's hard to distinguish between flux from the source and the flux from the background, and no correction is applied. We tried to minimize the effects of imperfect background subtraction by selecting bright objects (with smaller photometric error) for the fit. }
    \label{fig:curve_of_growth}
\end{figure*}

Only objects with relatively high S/N must be used for calculating the curve of growth to avoid background contamination, so we require the MAG\_AUTO error be smaller than these limits: [0.02, 0.01, 0.01, 0.01, 0.01, 0.01] for $u, g, r, i, i2, z$ in the CFHTLS Deep field, [0.05, 0.05, 0.05, 0.04, 0.05] for $u, g, r, i, z$ in CFHTLS Wide fields, 0.02 for the $Y$-band deep pointing and 0.05 for $Y$-band shallow pointing. We also exclude saturated, masked or blended objects by requiring the CFHTLS ``flag'' value to be $\leq 1$ and the SExtractor flag (in $r$-band or $Y$-band) to be 0. Fig. \ref{fig:curve_of_growth} shows examples of the curve of growth fits.

We then obtain $\alpha$ and $\beta$ by fitting equation \ref{eq:fraction} to the measured curve of growth for a given radius bin by least squares.
Once we know $\alpha$ and $\beta$, we can measure the flux of each object in a small aperture $r_0$, and extrapolate to infinity to obtain the total flux. Essentially, we have then determined the aperture correction factor for a given radius bin:
\begin{equation}
\label{eq:correction_factor}
\mathrm{ApCorr} = \mathrm{frac}(\infty)/\mathrm{frac}(r_0) = 
\left\{\frac{\beta-1}{\pi \alpha^2}\left[1+\left(\frac{r_0}{\alpha}\right)^2\right]^{-\beta}\right\}^{-1},
\end{equation}
where $\alpha$ and $\beta$ are fit separately for each  bin.

For the $ugriz$ bands, we choose the aperture radius $r_0 = 5 $ pixels ($0.93\arcsec$), because among available apertures this choice yielded the highest signal-to-noise photometry for all but the brightest objects. For $Y$-band we choose a similar aperture radius of $r_0 = 4.5 $ pixels ($0.9\arcsec$).

\begin{figure*}
    \centering
    \begin{subfigure}[b]{0.33\textwidth}
        \includegraphics[width=\textwidth]{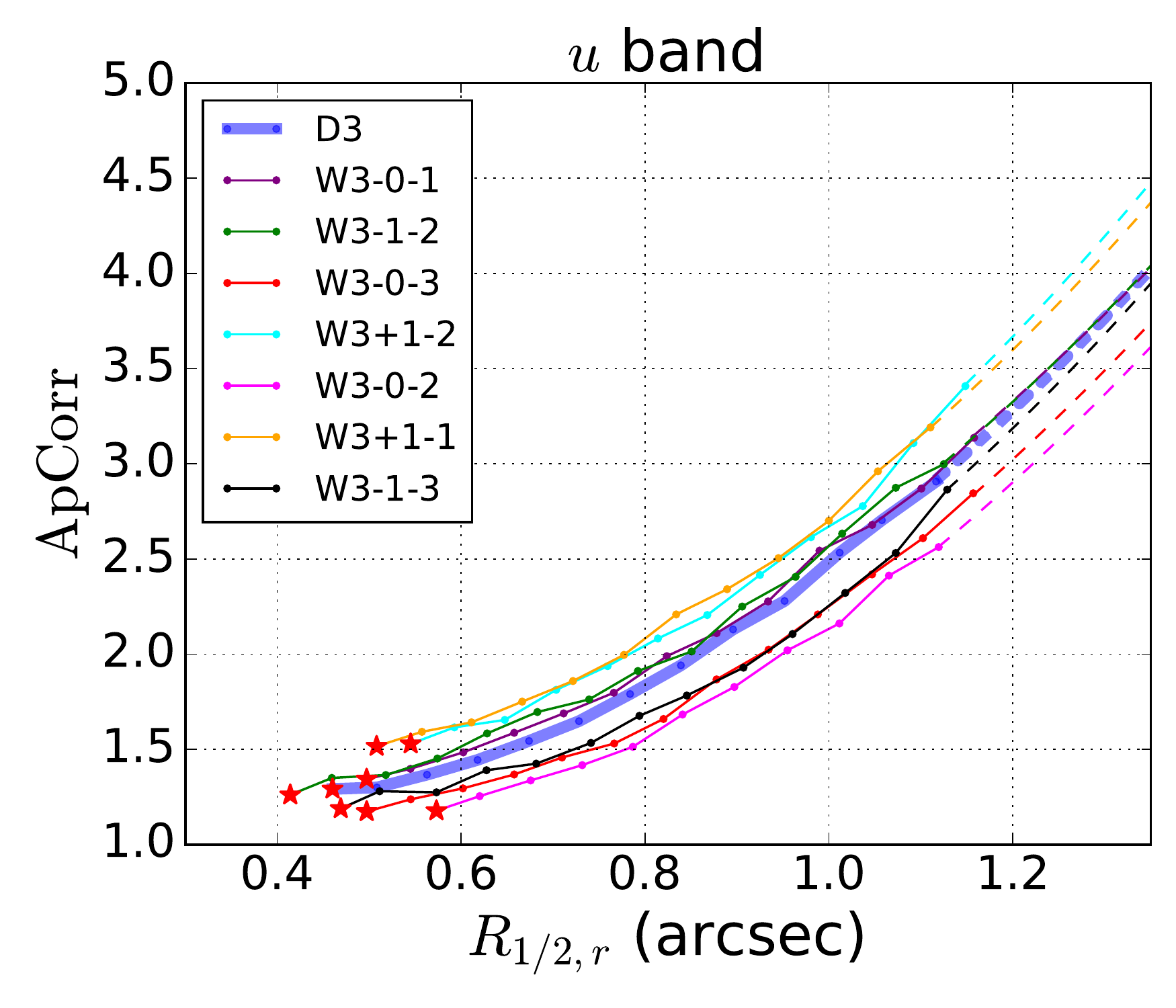}
        \caption{}
    \end{subfigure}
    \begin{subfigure}[b]{0.33\textwidth}
        \includegraphics[width=\textwidth]{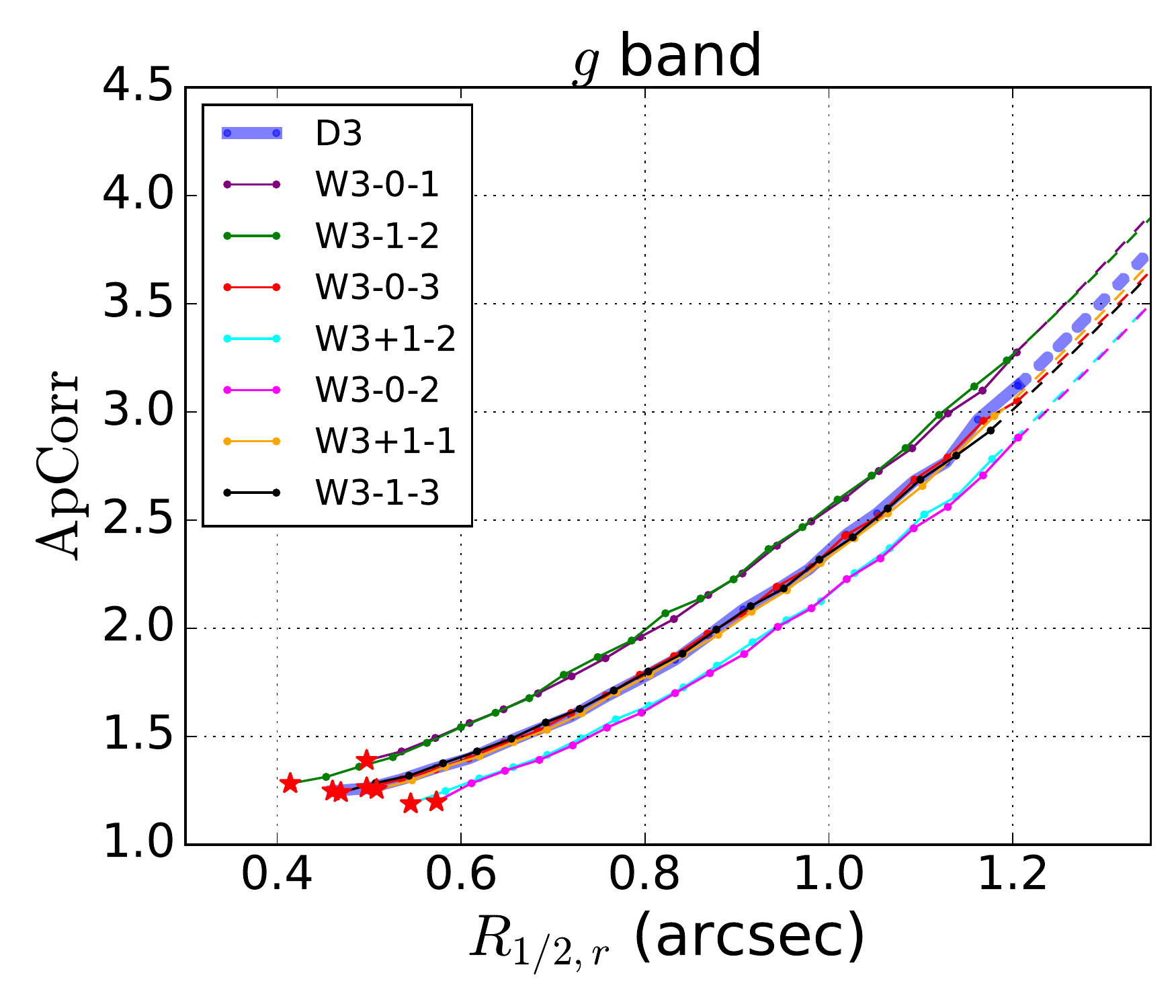}
        \caption{}
    \end{subfigure}
    \begin{subfigure}[b]{0.33\textwidth}
        \includegraphics[width=\textwidth]{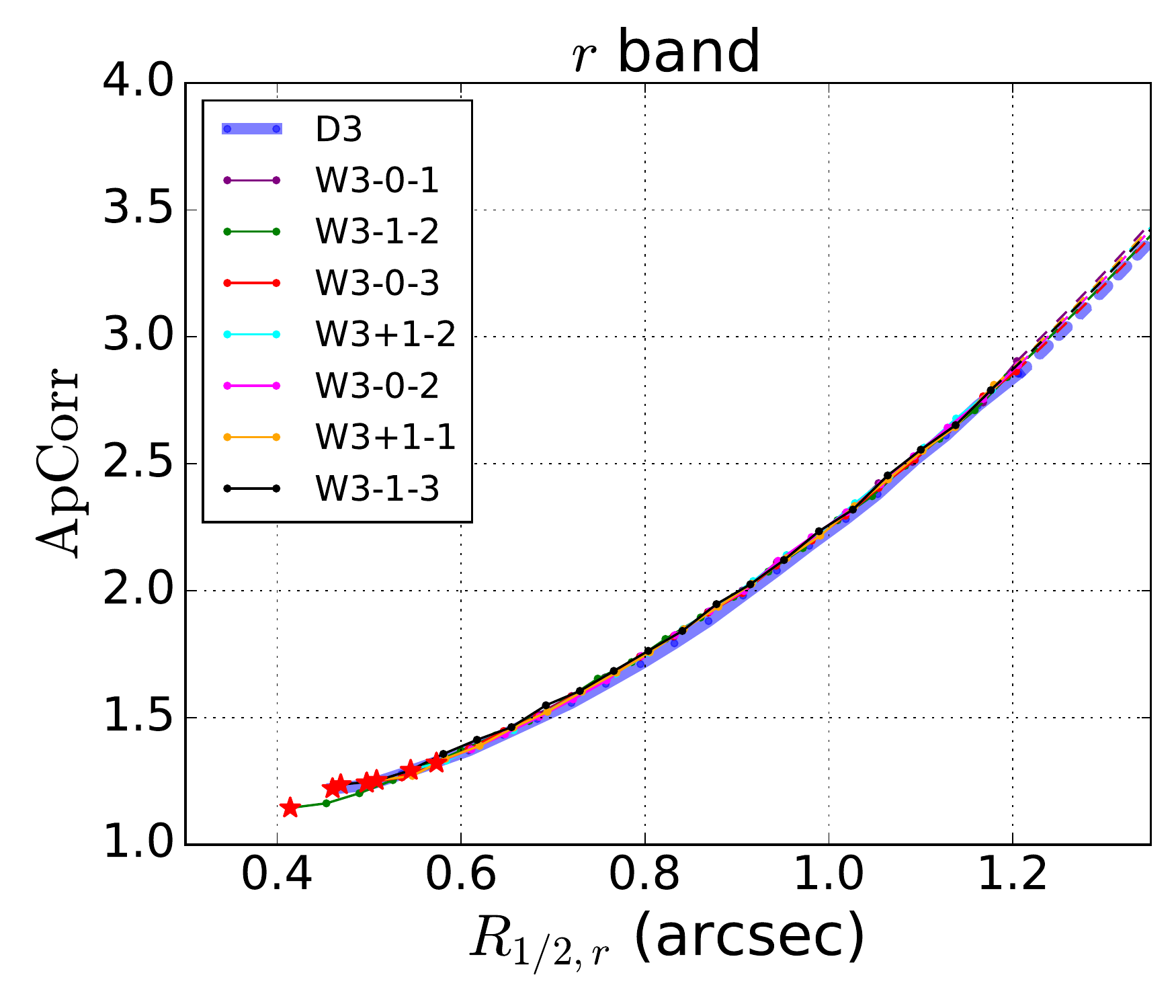}
        \caption{}
    \end{subfigure}\\
    \begin{subfigure}[b]{0.33\textwidth}
        \includegraphics[width=\textwidth]{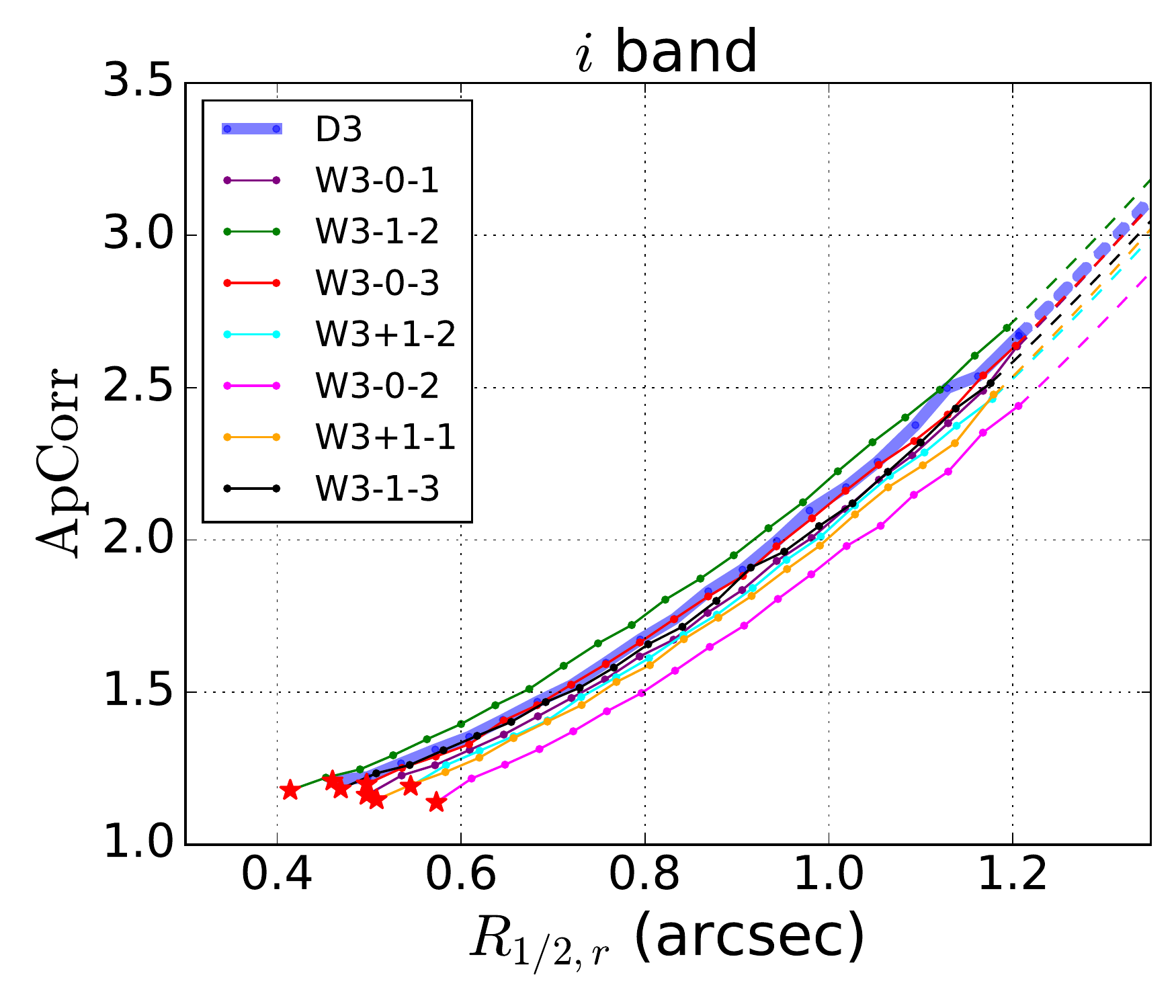}
        \caption{}
    \end{subfigure}
    \begin{subfigure}[b]{0.33\textwidth}
        \includegraphics[width=\textwidth]{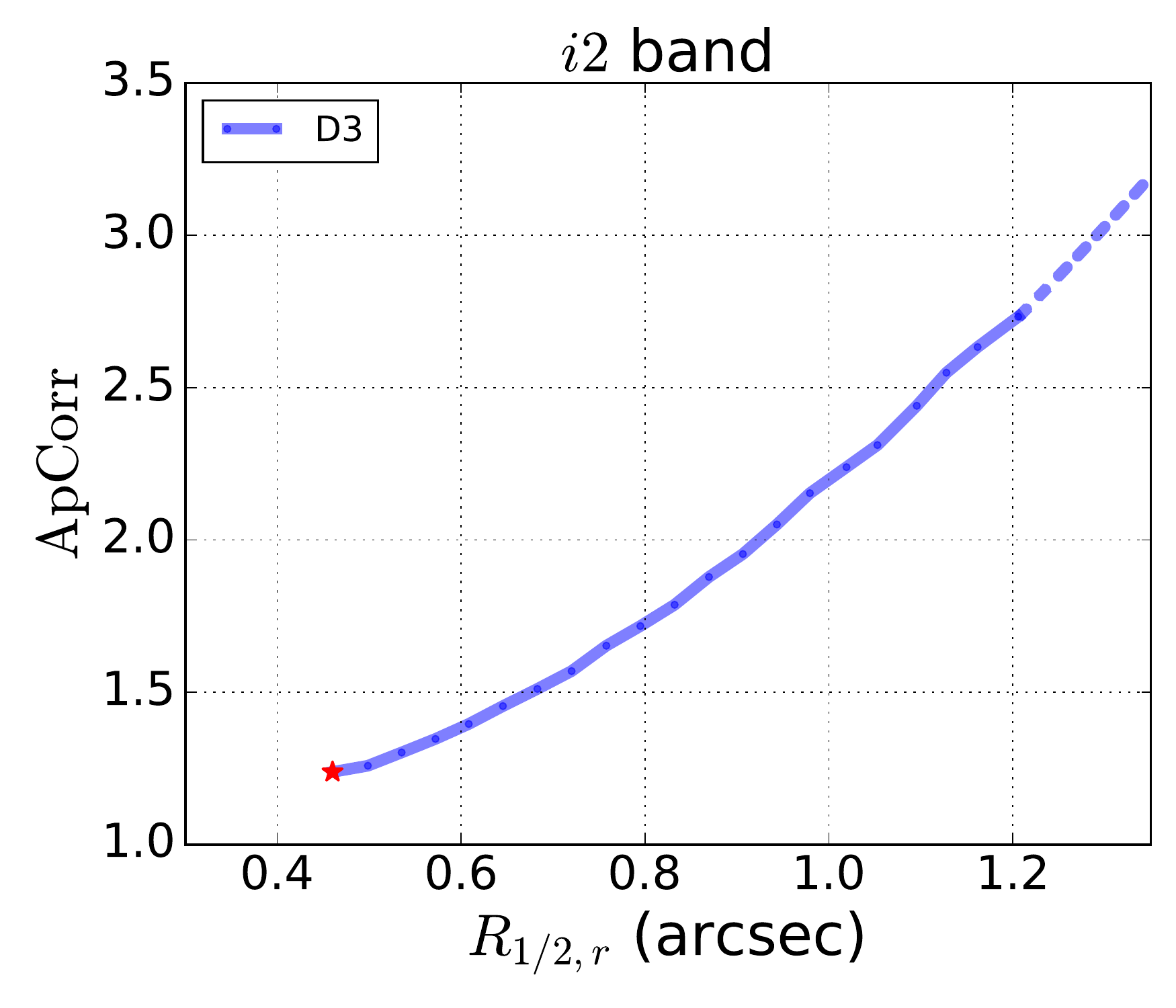}
        \caption{}
    \end{subfigure}
    \begin{subfigure}[b]{0.33\textwidth}
        \includegraphics[width=\textwidth]{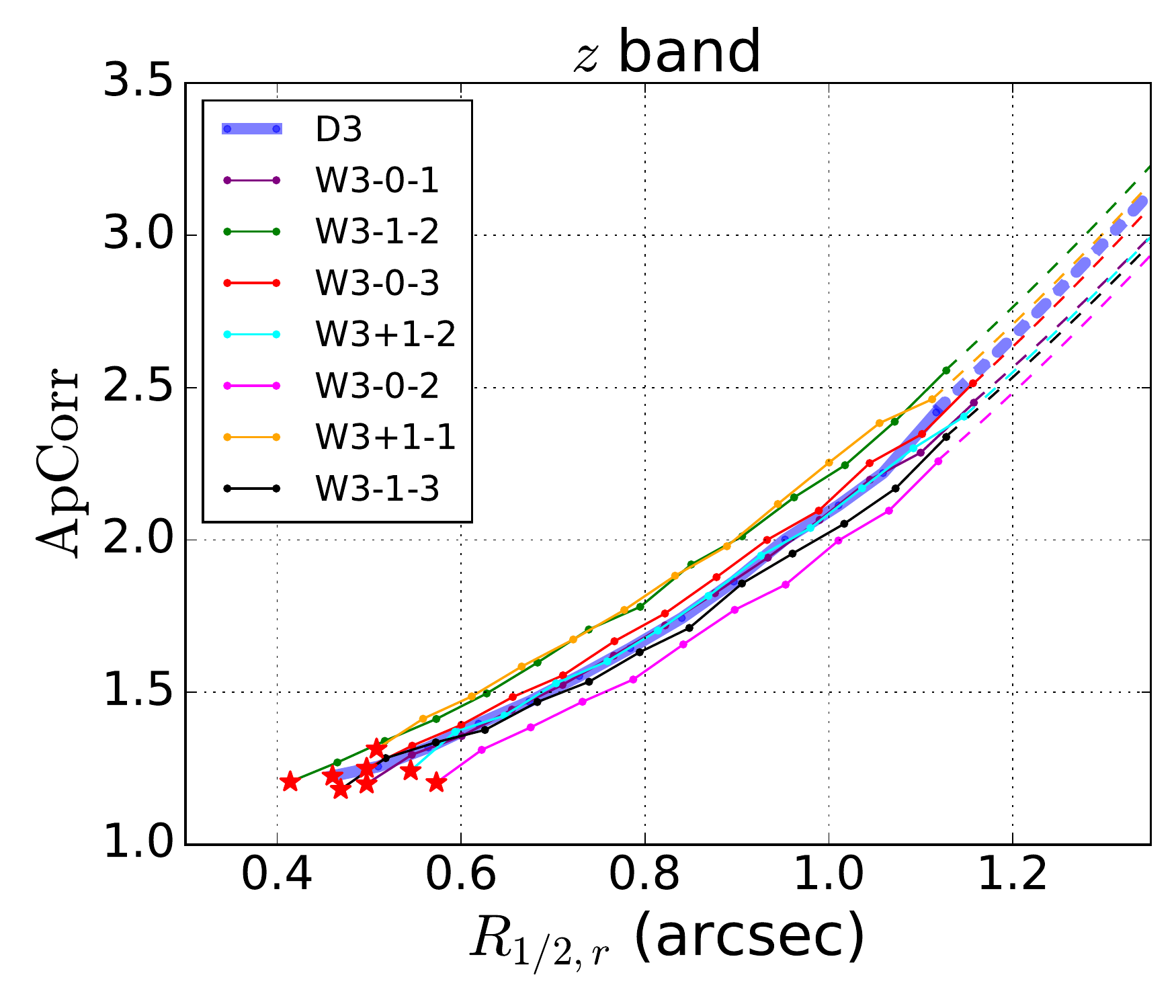}
        \caption{}
    \end{subfigure}\\
    \begin{subfigure}[b]{0.33\textwidth}
        \includegraphics[width=\textwidth]{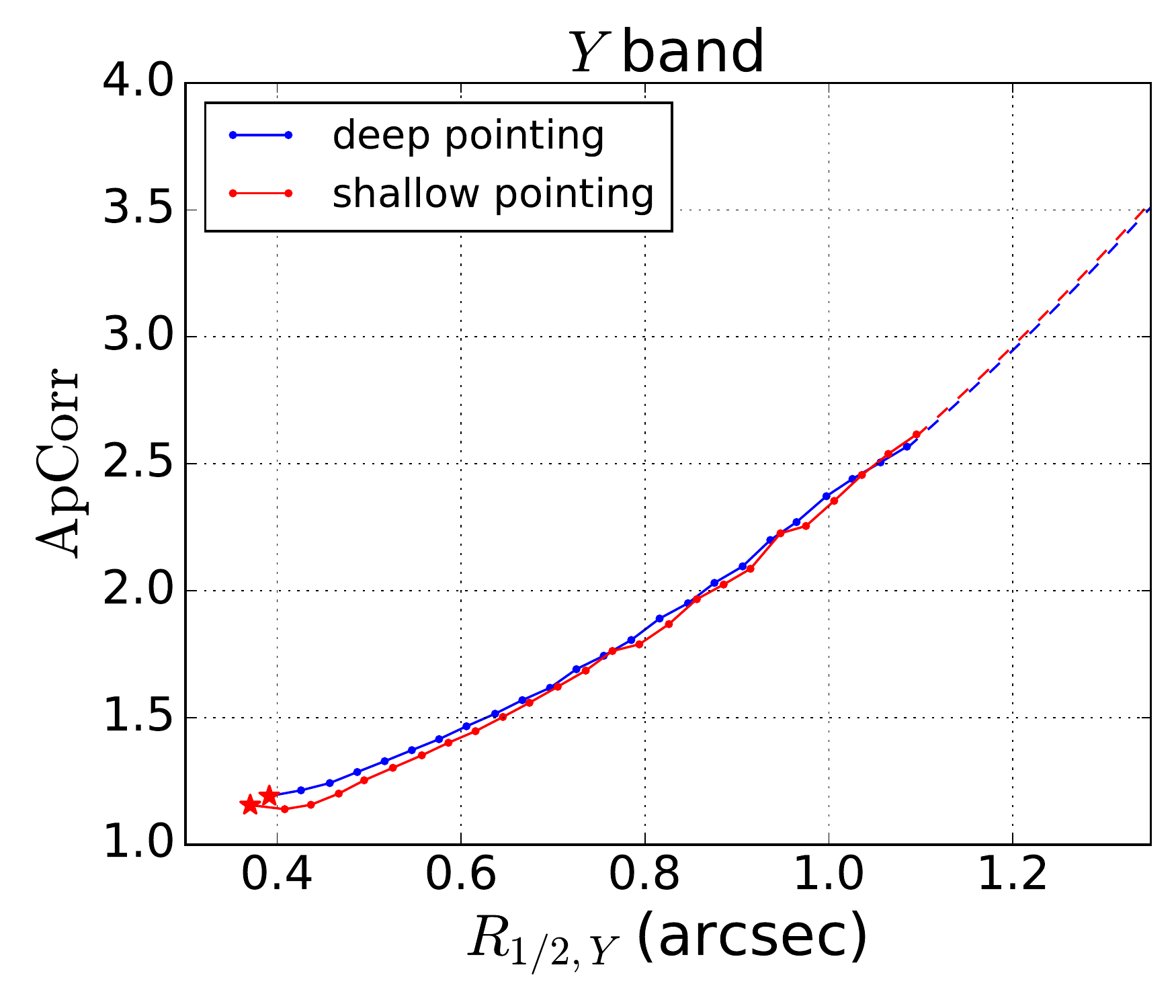}
        \caption{}
    \end{subfigure}
\caption{Each plot shows the correction factor ApCorr in one band as a function of half-light radius, over-plotting all pointings . In (a-f) the thick line is the Deep field D3 and the thin lines are the Wide field W3 pointings. (g) shows the $Y$-band. The correction factor and radius of stars are plotted as the star marker. The dashed line is the extrapolation for objects larger than the radius bins. }
\label{fig:flux_ratio}
\end{figure*}

After obtaining ApCorr for each radius bin, we calculate ApCorr as a function of radius by linear interpolation to determine the correction for each individual object. To obtain the correction factor for objects larger than the largest radius bins, we must extrapolate $\mathrm{ApCorr}(R_{1/2})$ to larger radii. To do this, we use the $\alpha$ and $\beta$ from the largest radius bin to calculate the actual fraction of light within the SExtractor ``half-light'' radius, and assume that this fraction is the same for all objects of larger radii; we then keep $\beta$ fixed and use the SExtractor ``half-light'' radius to estimate $\alpha$ and obtain ApCorr. Fig. \ref{fig:flux_ratio} shows the correction factor ApCorr as a function of radius. 

Finally, we use the function $\mathrm{ApCorr}(R_{1/2})$ to obtain the total flux from the aperture flux within aperture radius $r_0$ for every object in the catalog.

\subsection{Error estimation}
Assuming that our model of the star and galaxy light profiles is correct, the corrected aperture magnitude MAG\_APERCOR should have two sources of error: photometric errors in the aperture magnitude which were measured by SExtractor, and the error in the correction factor ApCorr which we multiplied by. In the catalog and in this paper, we label MAGERR\_APER (uerr\_aper, gerr\_aper, etc.) as the photometric error from SExtractor, and MAGERR\_APERCOR (uerr\_apercor, gerr\_apercor, etc.) as the statistical uncertainty in the correction factor.

Here we assume that the error in $\mathrm{ApCorr}(R_{1/2})$ is only due to the error in the radius $R_{1/2}$, and the correction factor itself has negligible error if the radius is accurate. SExtractor does not provide the error in the radius, so we can only estimate this quantity indirectly. For $ugriz$ bands where the $r$-band radius is used, we assume that the $i$-band radius error $\sigma_{R_i}$ is the same as the $r$-band radius error $\sigma_{R_r}$, and since they are independent measurements, we can estimate $\sigma_{R_r}$ from the scatter of $f_{i,r} = R_i/R_r$ about its mean value, so that
\begin{equation}
    \frac{\sigma_{R_r}}{R_r} = \frac{\sigma_{f_{i,r}}}{\sqrt{2}\bar{f}_{i,r}}.
\end{equation}
Here $\bar{f}_{i,r}$ in the denominator is the average value of $f_{i,r}$. The radius error increases with decreasing S/N, so we calculate $\sigma_{f_{i,r}}$ for objects in $r$-band magnitude bins, and we obtained the fractional radius error $\sigma_{R_r}/R_r$ as a function of magnitude. Similarly, we can assume that $\sigma_{R_g}=\sigma_{R_r}$, and calculate $\sigma_{R_r}/R_r$ using $f_{g,r} = R_g/R_r$. We find that the fractional radius errors from $g$-band and $i$-band are consistent, and therefore we simply use the average of the two results as the final fractional radius error. Given the resulting estimate of the fractional radius error, we calculate MAGERR\_APERCOR for each object via propagation of errors:
\begin{equation}
    \mathrm{MAGERR\_APERCOR} = \frac{\sigma_{A}}{A} = \frac{1}{A} \frac{dA}{dR_r}R_r\frac{\sigma_{R_r}}{R_r},
\end{equation}
where $A$ is short for ApCorr. Similarly, in the $Y$-band, we match the objects to CFHTLS, and estimate $\sigma_{R_Y}$ and MAGERR\_APERCOR from the scatter of $f_{z,Y} = R_z/R_Y$.

In cases where one wishes to estimate the uncertainty in the total magnitude of an objects, the net error in MAG\_APERCOR is
\begin{equation}
\resizebox{\columnwidth}{!}{$\sigma_{\mathrm{MAG\_APERCOR}} = \sqrt{(\mathrm{MAGERR\_APER})^2 + (\mathrm{MAGERR\_APERCOR})^2}.$}
\end{equation}

Since the $r$-band radius is used for aperture correction for all of $ugriz$, the correction error MAGERR\_APERCOR is correlated and mostly cancels out when we calculate colors involving the $ugriz$ bands. For example, the error in $u-g$ color is
\begin{equation}
\resizebox{\columnwidth}{!}{$\sigma_{u-g} = \sqrt{\mathrm{UERR\_APER}^2 + \mathrm{GERR\_APER}^2 + (\mathrm{UERR\_APERCOR }-\mathrm{GERR\_APERCOR})^2}.$}
\end{equation}

The $Y$-band aperture correction did not use $r$-band radius, and the error in $z-Y$ is
\begin{equation}
\resizebox{\columnwidth}{!}{$\sigma_{z-y} = \sqrt{\mathrm{ZERR\_APER}^2 + \mathrm{YERR\_APER}^2 + \mathrm{ZERR\_APERCOR}^2 + \mathrm{YERR\_APERCOR}^2}.$}
\end{equation}

Similar formulae may be used to determine the net uncertainty in any color derived from these passbands.


\bibliographystyle{mnras}
\bibliography{LSST_testbed}


\bsp    
\label{lastpage}
\end{document}